\documentclass[fleqn,usenatbib]{mnras}

\usepackage{newtxtext,newtxmath}
\usepackage{booktabs}
\usepackage{longtable}
\usepackage[dvipsnames]{xcolor}

\usepackage[T1]{fontenc}

\DeclareRobustCommand{\VAN}[3]{#2}
\let\VANthebibliography\thebibliography
\def\thebibliography{\DeclareRobustCommand{\VAN}[3]{##3}\VANthebibliography}

\usepackage{graphicx}	
\usepackage{amsmath}	
\usepackage{ulem}

\newcommand{\totalMSsources}{1,147,017}
\newcommand{\nNonClassfied}{31,550} 
\newcommand{\nQSOCandidates}{11,612}
\newcommand{\nStarCandidates}{19,198}
\newcommand{\nGalCandidates}{732}
\newcommand{\nQSOCandidatesColors}{11,696}
\newcommand{\nStarCandidatesColors}{19,191}
\newcommand{\nGalCandidatesColors}{663}
\newcommand{\candMag}{1,028}
\newcommand{\candCol}{898}

\title[Synthetic data applied to the QUBRICS selection]{The probabilistic random forest applied to the QUBRICS survey: improving the selection of high-redshift quasars with synthetic data}

\author[Guarneri  et al.]{
\parbox[t]{\textwidth}{
Francesco Guarneri$^{1,2}$\thanks{E-mail: francesco.guarneri@inaf.it},
Giorgio Calderone$^{2}$,
Stefano Cristiani$^{2,3,4}$,
Matteo Porru$^{1}$,
Fabio Fontanot$^{2}$,
Konstantina Boutsia$^{5}$,
Guido Cupani$^{2,3}$, 
Andrea Grazian$^{6}$,
Valentina D'Odorico$^{2,3,7}$,
Michael T. Murphy$^{3, 8}$,
Angela Bongiorno$^{9}$,
Ivano Saccheo$^{9}$,
Luciano Nicastro$^{10}$
}
\vspace*{6pt}\\
$^{1}$ Dipartimento di Fisica, Sezione di Astronomia, Università di Trieste, via G.B. Tiepolo 11, I-34131, Trieste, Italy \\
$^{2}$ INAF--Osservatorio Astronomico di Trieste, Via G.B. Tiepolo, 11, I-34143 Trieste, Italy \\
$^3$ IFPU--Institute for Fundamental Physics of the Universe, via Beirut 2, I-34151 Trieste, Italy \\
$^4$ INFN--National Institute for Nuclear Physics,  
via Valerio 2, I-34127 Trieste, Italy \\
$^5$ Las Campanas Observatory, Carnegie Observatories, 
Colina El Pino, Casilla 601, La Serena, Chile\\
$^6$ INAF--Osservatorio Astronomico di Padova,
Vicolo dell'Osservatorio 5, I-35122 Padova, Italy \\
$^7$ Scuola Normale Superiore, P.zza dei Cavalieri, I-56126 Pisa, Italy\\
$^8$ Centre for Astrophysics and Supercomputing, Swinburne University of Technology, Hawthorn, Victoria 3122, Australia \\
$^9$ INAF--Osservatorio Astronomico di Roma, Via Frascati 33, I-00078 Monte Porzio Catone, Italy \\
$^{10}$ INAF--Osservatorio di Astrofisica e Scienza dello Spazio di Bologna, Via P. Gobetti 101, I-40129 Bologna, Italy
}

\date{Accepted XXX. Received YYY; in original form ZZZ}

\pubyear{2022}

\begin{document}
\label{firstpage}
\pagerange{\pageref{firstpage}--\pageref{lastpage}}
\maketitle

\begin{abstract}
Several recent works have focused on the search for bright, high-$z$ quasars (QSOs) in the South. Among them, the QUasars as BRIght beacons for Cosmology in the Southern hemisphere (QUBRICS) survey has now delivered hundreds of new spectroscopically confirmed QSOs selected by means of machine learning algorithms.
Building upon the results obtained by introducing the probabilistic random forest (PRF) for the QUBRICS selection, we explore in this work the feasibility of training the algorithm on synthetic data to improve the completeness in the higher redshift bins. We also compare the performances of the algorithm if colours are used as primary features instead of magnitudes.
We generate synthetic data based on a composite QSO spectral energy distribution. We first train the PRF to identify QSOs among stars and galaxies, then separate high-$z$ quasar from low-$z$ contaminants. We apply the algorithm on an updated dataset, based on SkyMapper DR3, combined with Gaia eDR3, 2MASS and WISE magnitudes. We find that employing colours as features slightly improves the results with respect to the algorithm trained on magnitude data. Adding synthetic data to the training set provides significantly better results with respect to the PRF trained only on spectroscopically confirmed QSOs. We estimate, on a testing dataset, a completeness of $\sim86\%$ and a contamination of $\sim36\%$. Finally, 206 PRF-selected candidates were observed: 149/206 turned out to be genuine QSOs with z$ > 2.5$, 41 with z$ < 2.5$, 3 galaxies and 13 stars. The result confirms the ability of the PRF to select high-$z$ quasars in large datasets.
\end{abstract}




\section{Introduction}
Luminous, high-redshift quasars are of paramount importance for a wide range of extragalactic studies. These include, e.g., the number density of bright quasars at high-$z$ \citep{SDSSIncomplete_Schindler_PS:2019ApJ...871..258S,LF_Boutsia:2021ApJ...912..111B, Onken21:2021arXiv210512215O, LF_Hz_Grazian:2021arXiv211013736G}, their role in the re-ionisation process \citep[e.g.,][Fontanot et al., in preparation]{FontanotReionization12}, the early phases of galaxy formation and co-evolution with their central SMBHs \citep[e.g.,][]{FontanotSMBHEvo:2020MNRAS.496.3943F}, the inference of cosmological parameters from lensed QSOs or the characterisation of the accretion properties for SMBHs \citep[e.g.,][]{2015Natur.518..512W}.

High-$z$, bright QSOs allow to investigate the properties of the intergalactic medium, which at $z>1.5$ contains more than 80\% of all baryonic matter: light from these beacons carries information on the spatial distribution, chemical enrichment and kinetic properties of the interposed gas. Example of relevant studies include: the determination of primordial element abundances \citep[e.g., deuterium,][]{CookePettini14Deuterium, CookePettini17Deuterium}, the variation of fundamental constants \citep[e.g., the fine structure constant;][]{FundConstMilakovic:2021MNRAS.500....1M, FundConstMurphy:ApJ} or the ability to directly probe the expansion history of the Universe by measuring the redshfit drift  \citep[][]{Liske+08:2008MNRAS.386.1192L}.

The best candidates for these studies (i.e., bright, high-redshift QSOs) are, however, difficult to select in wide surveys. This is evident when comparing their surface densities with that of other objects at the same apparent magnitude: for instance, less than 1\% of the objects in the sample used in \citet[][]{Boutsia20:2020ApJS..250...26B} were quasars with $i_\mathrm{psf}<18$ and $z>2.5$. Historically, the largest efforts in searching for QSO targets have been made in the Northern Hemisphere: the SDSS, which has now delivered more than $7.5\cdot10^5$ spectroscopically confirmed QSOs between redshift 0 and 7 \citep{LykeSDSS16q:2020ApJS..250....8L}, is the most prominent example of these endeavours.
Inferior investment in telescope time in the Southern Hemisphere has led to a less favourable situation: 
undiscovered targets are waiting to be found, by exploiting recent photometric catalogues targeting the southern sky \citep[SkyMapper, DESI and PanSTARRS,][]{SkyMapper3:2019PASA...36...33O, DESY3Gold, Panstarrs}.

\begin{figure}
    \centering
    \includegraphics[width=\columnwidth]{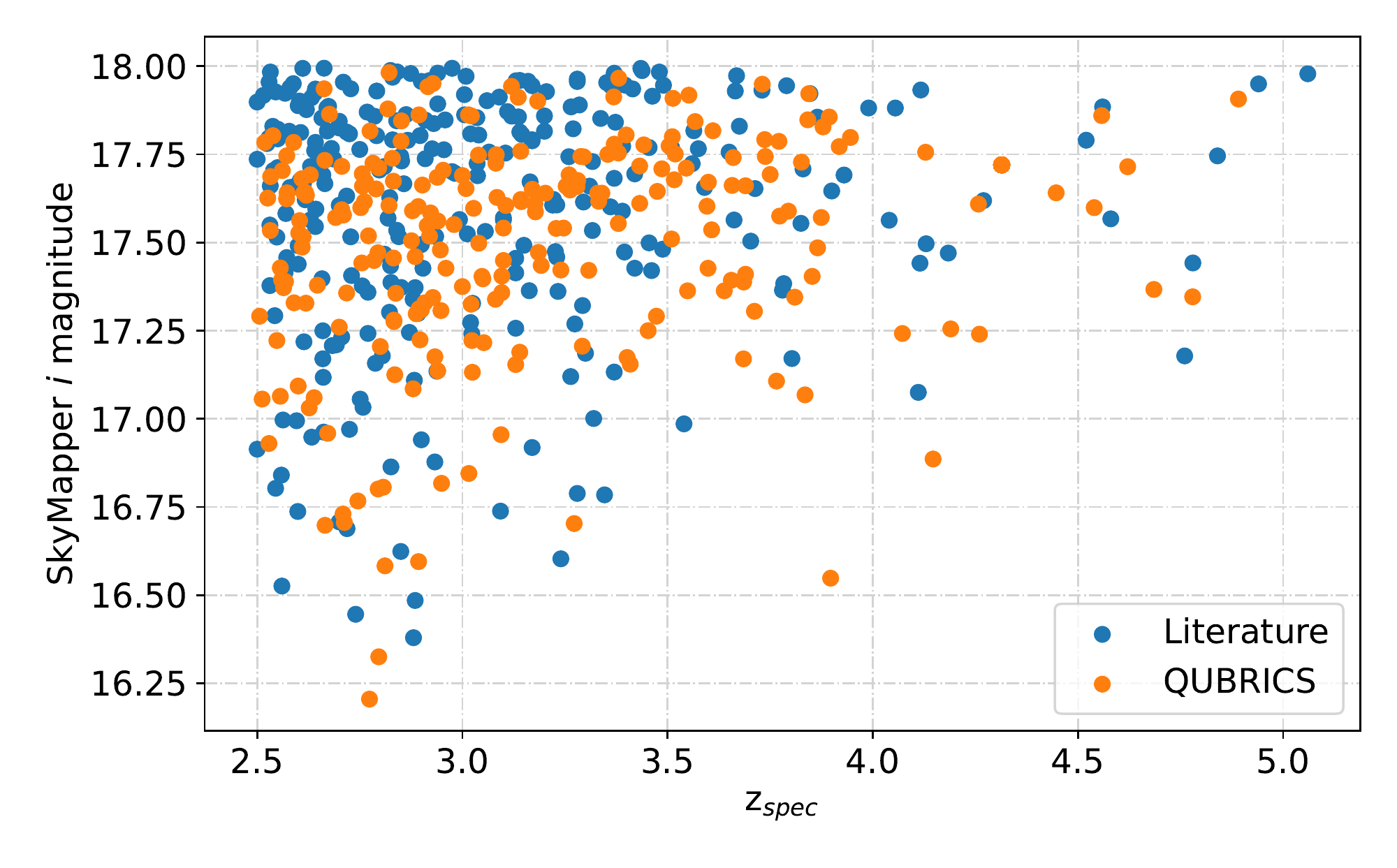}
    \caption{New QSOs identified by QUBRICS (orange dots), compared to QSOs spectroscopically confirmed taken from literature (blue dots). The plot shows all QSOs published until Paper I; new spectroscopic identifications have been obtained in the last year.}
    \label{fig:QSO_zspec_imag}
\end{figure}

Taking advantage of these large catalogues, the QUBRICS
program started to search for bright and high-$z$ QSOs by means of machine learning techniques applied on photometric data. The first catalogues of candidates have been derived using a canonical correlation analysis \citep[CCA, ][]{ref:CCA, Calderone19:2019ApJ...887..268C} trained on known, spectroscopically confirmed QSOs from literature. We applied the CCA both as classifier (i.e., an algorithm to label individual sources as star, non-active galaxy or QSO) and as regressor (i.e., as an estimator for the photometric redshift for each potential QSO). We obtained spectroscopical observations of the most promising quasar candidates with telescopes and confirmed the QSO nature of more than 250 sources \citep[Fig. \ref{fig:QSO_zspec_imag},][]{Calderone19:2019ApJ...887..268C, Boutsia20:2020ApJS..250...26B}.
In \citet[hereafter Paper I]{Guarneri:2021MNRAS.506.2471G}, we extended the range of available ML techniques by considering a new selection method based on the PRF.

The main advantages of this second method are the non linearity (and thus the ability to produce an improved predictive model), the ability to incorporate measurement uncertainties in the model, and to handle missing data. This paper builds upon the results presented in Paper I and reports our attempts in improving the sample selection of candidates (in particular at $z>3.5$). 

The paper is organised as follows: in section \ref{sec:catalogue_selectionAlg} we describe the base catalogue and the selection algorithm; in section \ref{sec:paperI} we review the approach followed in Paper I
and possible improvements. In particular, in order to obtain a
more effective training set we introduce synthetic QSOs. Section \ref{sec:synthDataGen} describes the method for generating synthetic data; section \ref{sec:testPerformance} and \ref{sec:candidateSample} concerns the analysis of the performance of the algorithm and the characterisation of the candidate sample; section \ref{sec:observations} reports the spectroscopic observations. Conclusions are finally drawn in section \ref{sec:conclusions}.

Unless stated otherwise, magnitudes are given in the AB magnitude system; uncertainties represent 68\% confidence intervals. We adopt a flat $\Lambda$CDM cosmology, with $\Omega_\mathrm{m}=0.3,\ \Omega_\Lambda = 0.7$, and $H_0=67.7\ \mathrm{kms^{-1}Mpc^{-1}}$.

\section{The selection methods and the catalogue} \label{sec:catalogue_selectionAlg}
\subsection{The selection algorithm}
Machine learning methods are becoming extremely common in astrophysics \citep[e.g.,][]{Baron19} to mine the growing wealth of information produced by modern astronomical facilities. The algorithm used for this work is the Probabilistic Random Forest \citep[][]{ReisPRF:2019AJ....157...16R}, a generalisation of the original random forest \citep[RF,][]{ref:BreimanRF} developed to account for measurements uncertainties. The PRF is applied as a supervised, classification algorithm, that maps features into discrete labels; in the current state, regression can not be performed with the PRF. In machine learning terms, features are properties associated to a given object in a dataset: in our case, features are magnitude estimates or colours (i.e. difference of magnitudes in two bands). Labels, instead, describe the class to which each object belongs to: in our case, we use star, non-active galaxies, low-$z$ QSO or high-$z$ QSO (with the latter two classes being separated at $z=2.5$).

We refer the interested reader to the original PRF paper \citep[][]{ReisPRF:2019AJ....157...16R} for a detailed description of the algorithm and the changes with respect to a classic random forest; we will provide below a very brief overview of the PRF algorithm.

Inheriting from the original random forest, the PRF is an ensemble algorithm that operates by creating a large number of decision trees. Each decision tree is a model described by a tree-like graph, built as a sequence of consecutive nodes. Intermediate nodes carry a condition, while the terminal node (i.e. a leaf of the tree) identifies a classification label; conditions and labels are determined during the training stage.
During the prediction stage, objects are propagated along each decision tree and classified according to the leaves they reach. Each tree provides an independent classification: a majority vote among all trees provides the final class returned by the forest. In addition to a class label, both the PRF and the RF output the probability, for a given object, of belonging to a given class. Both the random forest and the PRF map features to labels: the PRF, however, improves this mapping by taking into account measured uncertainties as well. This is accomplished by treating each input feature as a probability distribution function, with variance equal to the provided squared error and expectation value equal to the given feature. This change provides the PRF advantages over the original random forest: the model is more robust, with improved generalisation ability, while naturally handling missing data. This is achieved by propagating missing data to both sides of a node with equal probability. This last peculiarity is clearly desirable when working on photometric magnitudes, since missing data and non-detections are rather common.

\begin{figure}
    \centering
    \includegraphics[width=\columnwidth]{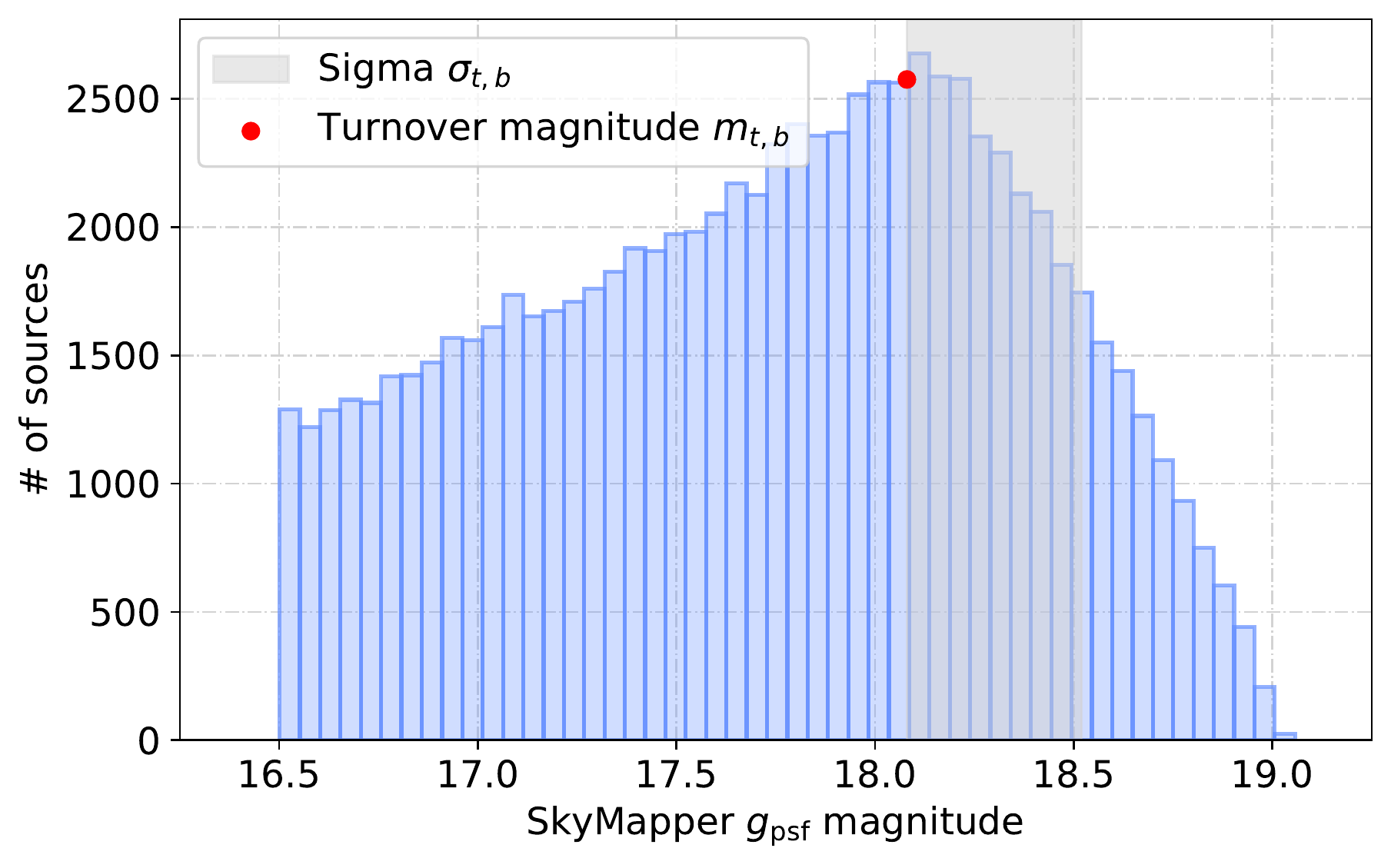}
    \caption{Magnitude distribution for a sample of randomly selected targets in the SkyMapper survey $g$ band.}
    \label{fig:LowPassDistribution}
\end{figure}

It is however necessary to distinguish between true missing data and non-detection. To account for this, the PRF code was modified to allow multiple distribution functions. Magnitude measurements were modelled as Gaussian distributions with mean equal to the feature value and variance equal to the provided error squared (as in the default version of the algorithm); upper limits were instead characterised by a low-pass distribution 
\begin{equation}
    \label{eq:uldistBase}
    f(x) = 1 - e^{-x} \hspace{1cm} x = \frac{x'-m_{t, b}}{\sigma_{t, b}}.
\end{equation} 
where $x'$ corresponds to an empirical approximation of the upper limit, $m_{t, b}$ to the turnover magnitude and $\sigma_{t,b}$ approximates the width of the faint-end of the count distribution (Fig. \ref{fig:LowPassDistribution}).

Following the same two-step approach described in Paper I, we adopted the Probabilistic Random Forest in order to select the most promising high-redshift QSO candidate.

In the first step we classify unlabelled sources as either QSOs, stars or galaxies based on their photometric properties. In the second, objects predicted to be QSOs are then reclassified as high- ($z > 2.5$) or low- ($z \leq 2.5$) redshift candidates. In both cases, we first train the PRF: as for the first step, we employ stars, galaxies and QSOs at all redshifts; we then use QSOs only as the training set for the second step. Once the algorithm is trained, it is applied on a validation/testing dataset to evaluate its performances. Instead of using all stars, we under-sample the available ones to uniformly cover the available $i-z$ colour space and match the number of available QSOs and galaxies, reducing their number from $\sim800,000$ to $\sim7,500$. For the second step, the algorithm is only trained on QSO data. Unfortunately, the number of high-redshift QSOs is $\sim1/8$ compared to the number of their low-z counterpart. To account for this, in Paper I high-redshift QSOs were over-sampled to provide a more uniform redshift distribution; in this work instead synthetic data are introduced as an alternative to oversampling. 

\subsection{The new Main Sample} \label{sec:NewMS}
The multi-wavelength photometric database covers the observer's frame from the UV to the far IR and was derived from the following surveys:
\begin{enumerate}
    \item $u_{\rm psf}$, $v_{\rm psf}$, $g_{\rm psf}$, $r_{\rm psf}$, $i_{\rm psf}$, $z_{\rm psf}$ magnitudes from the SkyMapper DR3 survey \citep{SkyMapper3:2019PASA...36...33O}
    \item $G$, $G_{BP}$, $G_{RP}$ magnitudes from the Gaia eDR3 catalogue \citep{GaiaEDr3:2021A&A...649A...1G}
    \item $J$, $H$, $K$ from 2MASS \citep{2MASS:2006AJ....131.1163S}
    \item W1, W2, W3, W4 from the AllWise \citep{WISE:2010AJ....140.1868W} survey.
\end{enumerate}
The choice of point spread function magnitudes is motivated by the point-like appearance of QSOs.

\begin{figure}
    \centering
    \includegraphics[width=\columnwidth]{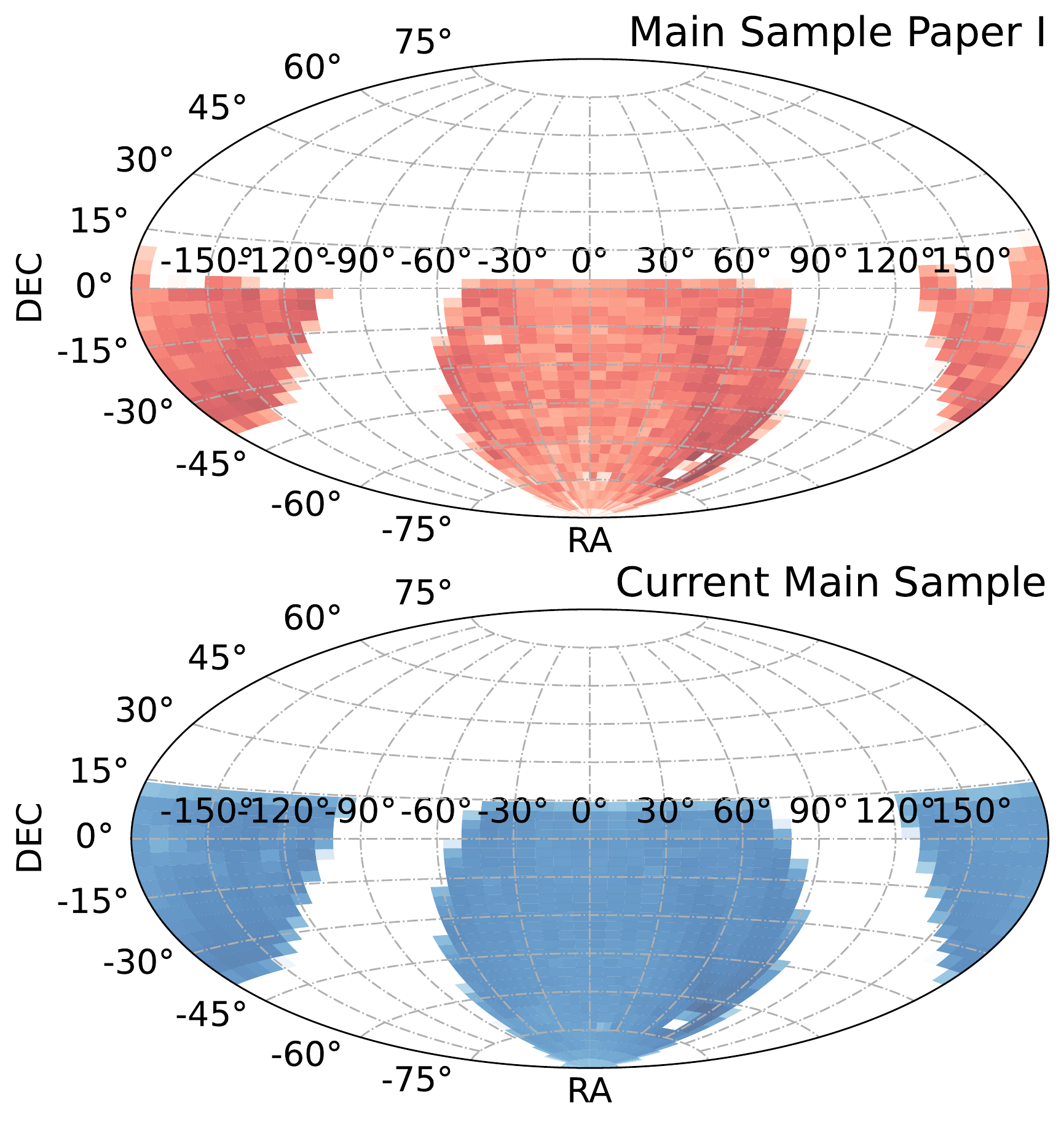}
    \caption{Current Main Sample compared to the Main Sample used in Paper I and based on SkyMapper DR1.1. The new sample is plotted in blue (lower panel) and is based on the third SkyMapper data release; darker shades of red reflect higher object density in the area. The Main Sample used in Paper I is plotted in the lower panel in blue: the third data release extends toward DEC $\leq 15$ deg.}
    \label{fig:SurveyArea}
\end{figure}

The main sample, hereafter MS, contains all the sources satisfying the following conditions:
\begin{itemize}
    \item $i_{\rm psf}$ magnitudes with photometric flag \texttt{i\_flags = 0} ($14 \leq i_{\rm psf} < 18$). This is equivalent to require that the $i_{psf}$ magnitude is free from warnings concerning saturation, close neighbours, edge-of-CCD effects and other systematic that might affect the photometry\footnote{For a complete list of possible systematic we refer the interested reader to the SkyMapper documentation (\url{https://skymapper.anu.edu.au/data-release/\#Coverage}).};
    \item $z_{psf}$ magnitudes with photometric flag \texttt{z\_flags = 0};
    \item Non \texttt{null} $G$ magnitudes from Gaia eDR3. Moreover, the associated source in the Gaia catalogue must be within 0.5'' of the SkyMapper detection;
    \item Non \texttt{null}, $SNR > 3$ AllWISE magnitudes in the W1, W2 and W3 bands. Moreover, the associated source in the AllWise catalogue must be within 0.5'' of the SkyMapper detection;
\end{itemize}
We restrict our selection to $14 \leq i_{\rm psf} < 18$, excluding crowded regions with galactic latitude $|b_{gal}| < 25$ deg: this includes \totalMSsources{} sources, spread on $\sim14,500 \rm ~deg^2$ mostly in the Southern Hemisphere.
The $G_{BP}$, $G_{RP}$ (Gaia eDR3), J, H, K (2MASS) and W4 (AllWISE) magnitudes were added to the database when available, but their existence was not a necessary condition for an object to be included in the MS.
It should be noted that the sample considered for this work is rather similar to the Main Sample employed in Paper I. The most significant differences are the exclusion of near and far UV magnitudes from GALEX \citep{BianchiGALEX:2020ApJS..250...36B} due to modelling uncertainties when generating synthetic data and a slightly larger area covered by the survey (Fig. \ref{fig:SurveyArea}), thanks to the third SkyMapper data release that includes regions around $0 < \mathrm{DEC} \lesssim15$ deg.

Objects with parallax and proper motion significantly different from zero ($>3\sigma$) were classified as stars (79\% of the MS) while those with high likelihood of being extended were identified by comparing Petrosian and PSF magnitudes \citep[17\% of the MS, see also][for more details]{Calderone19:2019ApJ...887..268C}. The remaining 4\% of sources were cross-matched against recent spectroscopic catalogues, e.g., the SDSSDR16Q \citep[][]{LykeSDSS16q:2020ApJS..250....8L}, the Veron-Cétty catalogue \citep{Veron10:2010A&A...518A..10V}, the 2dF \citep{2df:2001MNRAS.328.1039C}, the 6dF \citep{6df:2009MNRAS.399..683J}, in order to identify previously known QSOs and non-active galaxies. Additional spectroscopic identifications, when available, were drawn from the literature \citep[e.g.,][]{SDSSIncomplete_Schindler_PS:2019ApJ...871..258S,SDSSIncomplete_Schindler:2019ApJ...871..258S, Wolf20QSO:2020MNRAS.491.1970W, Onken21:2021arXiv210512215O} leaving \nNonClassfied{} unclassified sources: these will automatically be classified by the PRF trained on the spectroscopically confirmed QSOs and galaxies, together with $bona fide$ stars.

As described in \citet{Boutsia20:2020ApJS..250...26B}, the selection criteria adopted in this work might be biased against lensed sources, as positional matches were carried out with relatively stringent criteria to reduce contamination. This should not affect point-like sources, such as non-lensed QSOs, but might lead to the removal of extended objects (e.g., lensed sources with Einstein rings or crosses); moreover, the choice of some photometric bands (e.g., Gaia G) might remove very-high-redshift objects ($z \gtrsim 5$) from the initial sample. We decided against more relaxed constraints in order to maximise the success rate of telescope runs and maintain high success rate.

\section{Performance of the PRF in Paper I} \label{sec:paperI}
In Paper I we carried out a preliminary analysis on the performance of the PRF algorithm applied to the selection of high-redshift ($z>2.5$) QSOs. Extensive tests guided the choice of an approach aimed at maximising the success rate of the observations while still maintaining a sufficiently high completeness for applications such as the estimate of luminosity functions. The algorithm showed remarkably good performance in separating QSOs candidates from stars or non-active galaxies (especially for $z>2.5$ QSOs, as $\sim98\%$ of them were correctly classified as such).
In a second stage, high-redshift QSOs were sieved from lower-redshift QSOs: the estimated completeness, on a testing dataset, was $\sim84\%$, with a relatively low contamination of $\sim22\%$ of $z<2.5$ QSOs.

Spectroscopic follow-up and further tests 
allowed us to better characterise the properties of this PRF selection. 
A small number of candidates (16) turned out to be spectroscopically confirmed non-active galaxies or stars. In paper I we estimated their combined contribution to account for $\sim25$ of the 626 candidates found in the sample (4\% of the total); the current number of observed contaminants, with roughly half of the sample missing a spectroscopic confirmation, corroborates our estimate.
However, further tests have shown that few $z > 3$ QSOs were not successfully classified by the PRF. Misclassified sources are either among the higher redshift targets ($z\gtrsim4.6$) or the brighter, high-redshift QSOs, i.e., 
the incompleteness of the selection in Paper I affects the most precious QSOs for our scientific goals. 

We attempt to interpret the latter shortcoming as a consequence of non-optimal sampling of high-redshift QSOs in the training set: a larger, controlled dataset would improve the generalisation capabilities of the algorithm and is expected to recover at least part of the misclassifications. As mentioned in the introduction, the number of actual, spectroscopically confirmed, QSOs available to the scientific community in the Southern Hemisphere is still small, hence the training sample can not be improved using the data from the literature. Besides, failures (such as those listed in section \ref{sec:literatureMiscs}) of spectroscopic pipelines (e.g., SDSS) might affect the quality of the training dataset.

Synthetic data offer a solution to both problems: large samples can be generated rather easily, with complete control over the dataset features, and fed to the selection algorithm to improve its performance. 

\section{Synthetic data} \label{sec:synthDataGen}
Synthetic data are needed in particular at the highest redshifts
and at brighter magnitudes, where only a few QSOs are present in the 
training set. 
To produce synthetic spectral energy distributions (SEDs) we
rely on a relatively "standard" approach, parameterizing a composite QSO SED
and its expected variations, and extracting from it realizations
of the QUBRICS photometry for objects at the desired redshifts.

\begin{figure}
    \centering
    \includegraphics[width=8.75cm]{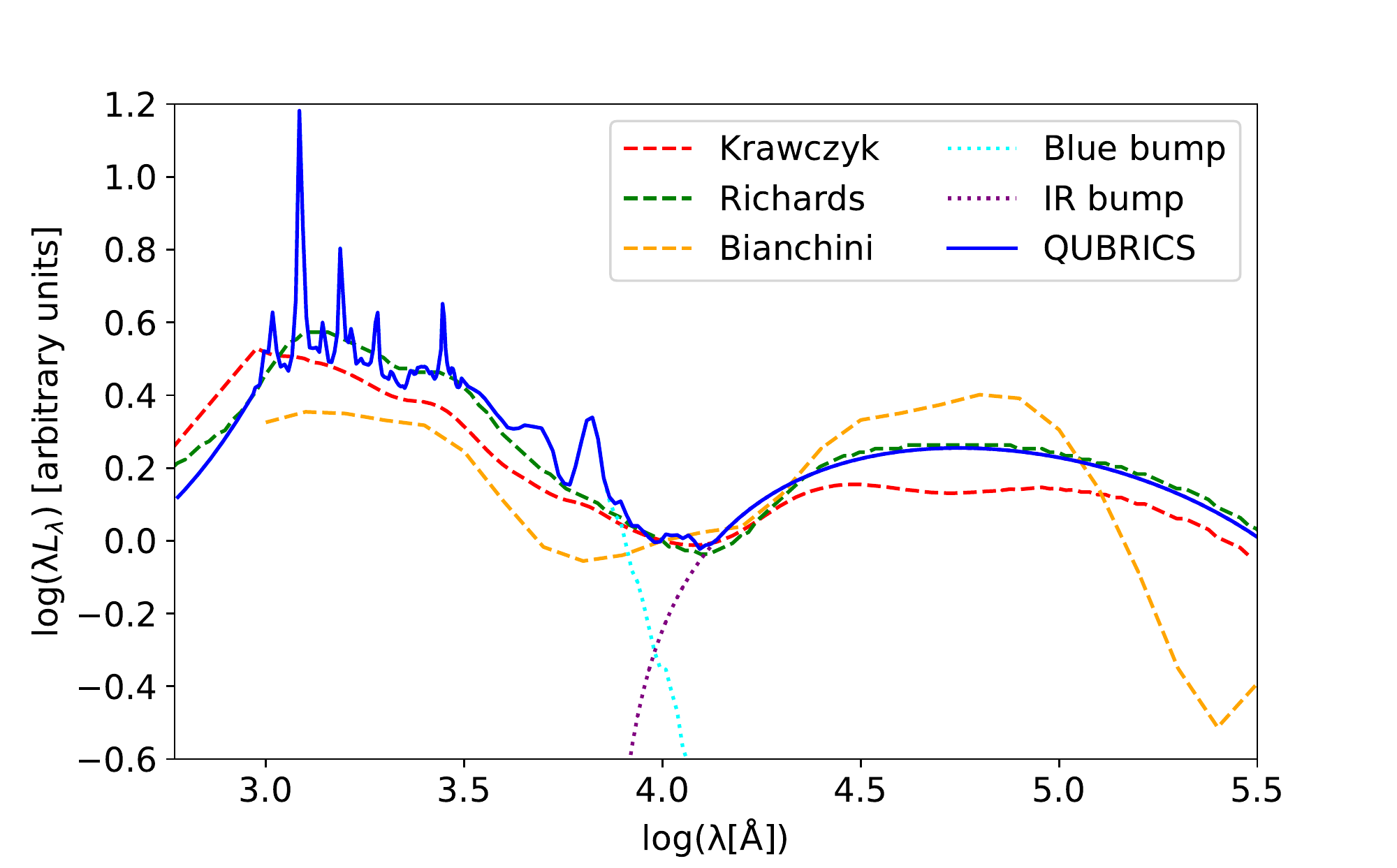}
    \caption{Composite spectral energy distribution of QSOs derived from the photometry of 6074 QSOs with known redshift (see text). The SED adopted in the present work is shown with a continuous blue line, with the dotted lines representing the two components (blue bump and IR bump) of the SED. For comparison the SEDs derived by \citet{Krawczyk2013}, \citet{Richards2006} and \citet{Bianchini2019} are shown with dashed lines. All the SEDs are normalised at $10^4$ \AA~ to a value $\log \lambda F_{\lambda} = 0.$ The continuous blue line corresponds to a SED obtained with a fractional contribution of the IR bump $f = 0.59$ and a slope $\alpha=0.0$ (see text).}
    \label{fig:QSO_SED}
\end{figure}

\subsection{The quasars' composite SED}
\label{sec:meanSED}
The first step for the generation of synthetic data consists in the parametrisation of the spectral energy distribution (SED) of quasars. We compiled a dataset with 6074 QSOs in the QUBRICS database with Gaia $G<18.5$
and spectroscopic redshift $z>0.5$. For each object, when available,
the photometry in the $G_{BP}$ and $G_{RB}$ bands, SkyMapper, 2MASS, WISE, SDSS-DR16, PanSTARRS1-DR2, VHS, GALEX was included.
For each of the 6074 QSOs we have parametrized its photometric data points in terms of: 

\begin{enumerate}
    \item a SED (shown in Fig~\ref{fig:QSO_SED} with a continuos blue line), adapted from the work by \cite{VandenBerk2001}, \cite{Richards2006} and \cite{Krawczyk2013}. We have empirically decomposed it in a "blue bump" (cyan dotted line), representing the accretion disk  \citep{Sun1989, Laor1989}, and an "IR bump" (purple dotted line), representing a dusty torus \citep{Pier1993, Mor2009}, with a cross-over at $10^4 \textup{~\AA}$.
    Different SEDs can be generated by chosing a different fractional contribution, $f$, of the "IR bump" with respect to the "blue bump";
    \item a multiplicative power-law, PL $ \propto (\lambda/2200 \textup{~\AA})^{- \alpha}$, applied to the "blue bump" part of the SED to account for extinction in the rest-frame optical-UV and possibly different spectral slopes in the blue bump;
    \item the spectroscopic redshift, used to compute the IGM absorption shortwards of 1215 $\textup{\AA}$, according to \citet{Inoue2014}, and shift the SED to the observer's frame;
\end{enumerate}

In this way a rest-frame SED is parametrized according to the following formula:
\begin{equation}
   SED(\lambda) = \frac{1-f}{0.41} \cdot F_A(\lambda) \cdot (\frac{\lambda}{2200})^{-\alpha} + \frac{f}{0.59} \cdot F_B(\lambda)
\end{equation}
where $F_A(\lambda)$ and $F_B(\lambda)$ are reported in Tab.~\ref{tab:SED_A} and Tab.~\ref{tab:SED_B}, respectively.

We obtain synthetic photometry by convolving the generated QSO SEDs, as would be measured in the observer's frame, with the transmission curves of the filters adopted in the input photometric surveys.
\begin{figure}
    \centering
    \includegraphics[width=8.75cm]{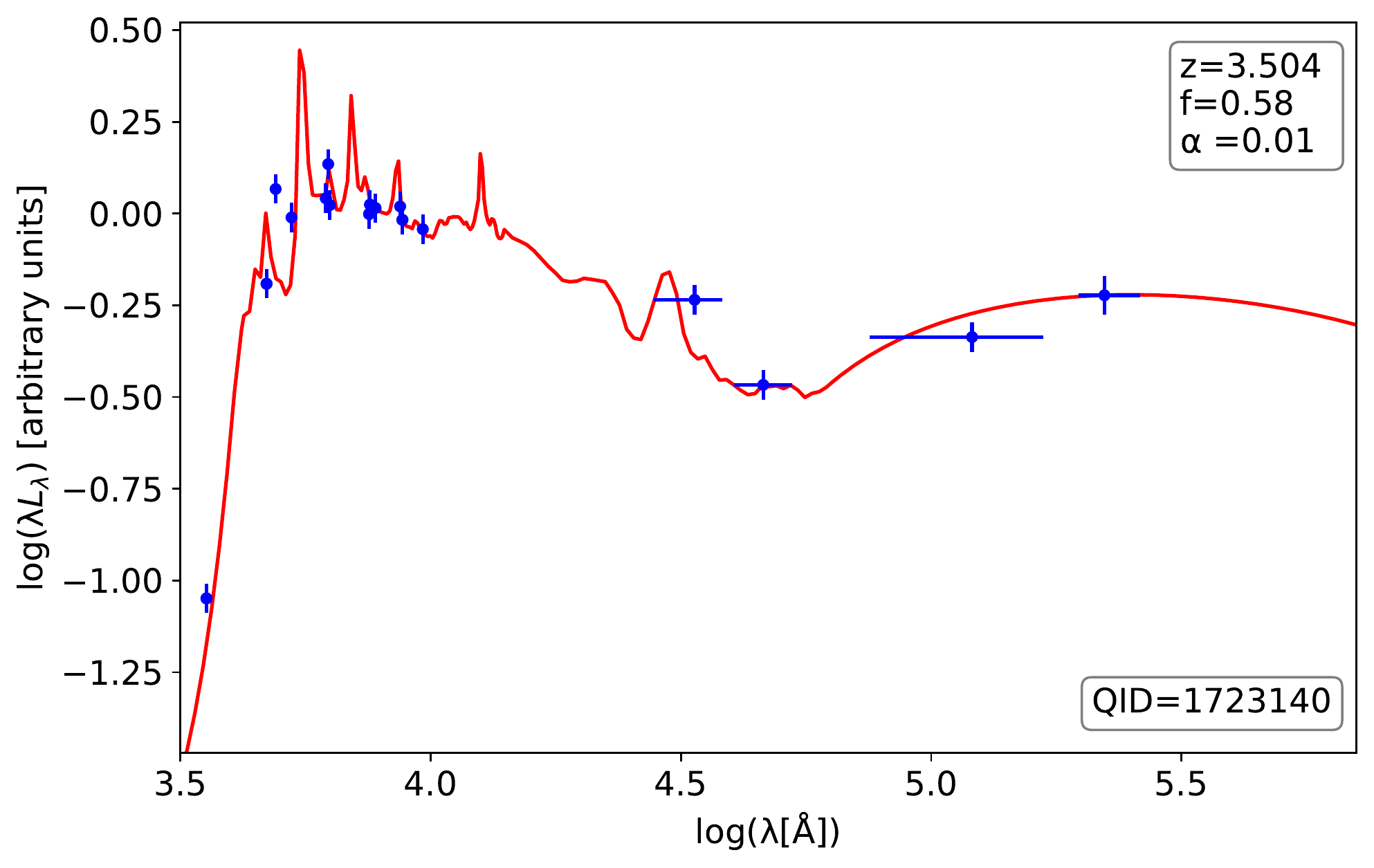}
    \caption{An example of SED fit for a QSO with spectroscopic redshift $z=3.504$. The parameterization of the SED is described in the text and the resulting optimal parameters, $f$ and $\alpha$, for the given photometric points are shown in the figure.}
    \label{Fig:SED_fit}
\end{figure}
For each of the 6074 known QSOs, given the spectroscopic redshift, we have fitted the fractional contribution, $f$, and the slope, $\alpha$,
optimising, chi-square-wise, the synthetic photometry with respect to the observed colours.
An example of SED fit for a QSO with spectroscopic redshift $z=3.504$ is shown in Fig.~\ref{Fig:SED_fit}. 

As a result, we obtain the distributions of $f$, whose median turns out to be $<f> = 0.57$, with a median absolute deviation (MAD) of 0.05 and of the slopes, $\alpha$,
with $<\alpha> = -0.01$ and MAD of 0.18.
In the next subsection these distributions will be used to generate synthetic photometric data of quasars at different redshifts, aiming at a fair representation of the QSO population.

\begin{table}
\centering
\begin{tabular}{|c|c|}
   \toprule
    Wavelength   & Relative Flux \\
    (\AA) & ({ \rm $F_{\lambda}$}) \\
    \midrule
600.0 & 0.431 \\
615.0 & 0.443 \\
630.0 & 0.456 \\
645.0 & 0.470 \\
660.0 & 0.483 \\
\dots & \dots \\
    \bottomrule
\end{tabular}
   \caption{A sample of the SED of the "blue bump" part of the composite QSO spectrum. Full version in electronic form.}
    \label{tab:SED_A}
\end{table}
\begin{table}
\centering
\begin{tabular}{|c|c|}
   \toprule
    Wavelength   & Relative Flux \\
    $\log$ (\AA) & ({ \rm $F_{\lambda}$}) \\
    \midrule
    7500.0 & 0.023 \\
    8000.0 & 0.065 \\
    8500.0 & 0.102 \\
    9000.0 & 0.136 \\
    9500.0 & 0.167 \\
    \dots & \dots \\
    \bottomrule
\end{tabular}
   \caption{A sample of the SED of the "IR bump" part of the composite QSO spectrum.
   Full version in electronic form.}
    \label{tab:SED_B}
\end{table}

\subsection{Generation of synthetic photometry}
\label{subsec:SynteticGeneration}
\begin{figure}
    \centering
    \includegraphics[width=\columnwidth]{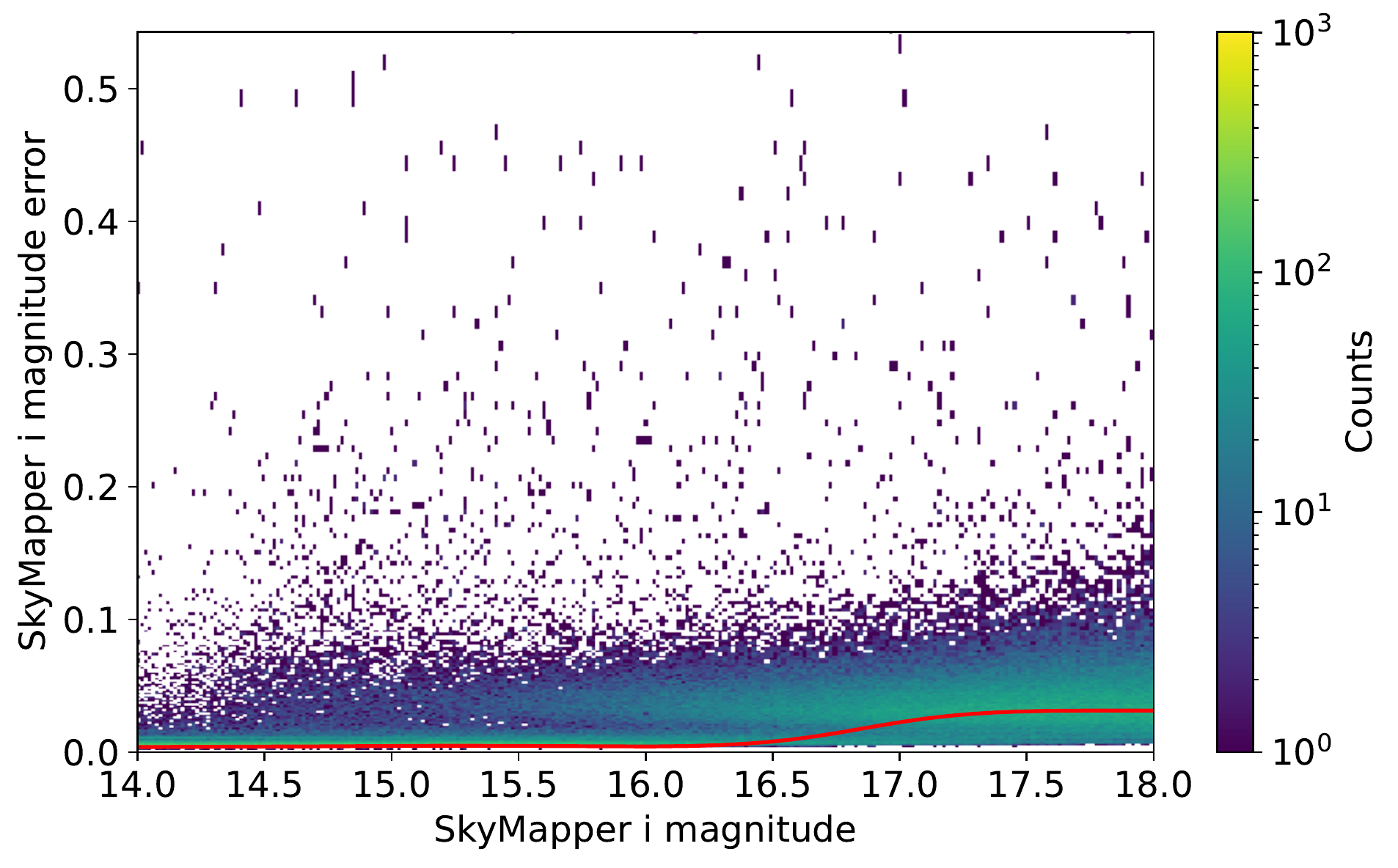}
    \caption{Example of a $f_e(m_b)$ (red line) computed for the SkyMapper $i$ band.}
    \label{fig:2DMagnitudeHistogram}
\end{figure}

In order to produce synthetic QSOs to be added to our training set
we proceeded as follows:
\begin{itemize}
    \item realisations of QSO SEDs were produced on the basis of the procedure described in the previous subsection, in particular extracting random values for the fractional contribution, $f$, of the IR bump with respect to the blue bump, and for the slope $\alpha$ from the observed distributions;
    \item the SEDs were redshifted in the observer's frame, and multiplied by the IGM transmission;
    \item the SEDs were multiplied with the surveys band-passes to obtain instrumental magnitudes; 
    \item the instrumental magnitudes were re-scaled in order to reproduce the distribution of the real sample. To this end, we considered again the sample of spectroscopically confirmed QSOs. We binned the available quasars in redshift: bins were chosen to ensure a minimum number of objects per bin ($\sim150$) and, when possible, a bin-width of $\sim0.2$.
    For each redshift bin, we considered the $i_{\rm psf}$ magnitude distribution and extracted a random number (using a Monte Carlo method) from the same distribution. We used the difference between the extracted number and the synthetic \textit{i} magnitude as offset for the magnitudes in all other bands; 
    \item we associated to each synthetic magnitude an error estimate. In order to do so, available photometric data were used to estimate an empirical error function $f_e(m_b)$. For each photometric band, we built a 2D histogram with magnitude on the x-axis and photometric error on the y-axis using all photometric data available in the MS (Fig. \ref{fig:2DMagnitudeHistogram}); the number of bins on both axis was chosen arbitrarily (250). For each magnitude bin (or, equivalently, each column in Fig. \ref{fig:2DMagnitudeHistogram}), we searched for the most populated error bin and interpolated, using a smoothing spline, each error point as a function of magnitude. This spline is the function $f_e(m_b)$, which returns the typical uncertainty assigned to the magnitude $m_b$. The scatter in the uncertainties as a function of magnitude, $s_e(m_b)$, was estimated as well using the same procedure. Rather than considering the most populated bin, we computed, as a first approximation for the scatter, the standard deviation across a column, and interpolated these points with a smoothing spline as a function of magnitude. \\ Given $f_e(m_b)$ and $s_e(m_b)$, the actual photometric uncertainty $e_{m_i}$ assigned to the magnitude $m_i$ was estimated by drawing a random value from a Gaussian with:
    
    \[ \begin{cases} 
      \mu = 0.1, \ \sigma = 0.01 & f_e(m_i) \leq 0.1 \\
      \mu = f_e(m_i),  \ \sigma = s_e(m_i) & f_e(m_i) > 0.1
    \end{cases}
    \]
    
    The process was repeated for each band in the catalogue;
    \item finally, we recomputed a noisy photometric magnitude. A new value was drawn from a Gaussian distribution with $\mu$ equal to the noiseless magnitude, derived from the synthetic SED, and $\sigma = e_{m_i}$. This is repeated for all magnitudes of a given object, and the newly found noisy magnitude, with the associated error, are added to the catalogue of synthetic objects.
\end{itemize}

In the catalogue of synthetic objects 
all sources have all the magnitude measurements and no missing data are present. Real photometric catalogues, however, often have missing data or non-detections, which are kept into account in the PRF selection, as described in Paper I and, briefly, in section \ref{sec:catalogue_selectionAlg}.

To mimic a more realistic situation, non-detections are introduced in the synthetic catalogue. We consider again Eq. \ref{eq:uldistBase}: given $m_{i, b}$, a magnitude measurement in the photometric band $b$, the turnover magnitude $m_{t, b}$ in the $b$ band and its associated error $\sigma_{t, b}$, the probability $P_{\rm ul}$ of $m_{i, b}$ to become an upper-limit is calculated using eq. \ref{eq:uldistBase} as
\begin{equation} \label{eq:uldist}
    P_{\rm ul} = f\left(\frac{m_{i, b} - m_{t, b}}{\sigma_{t, b}}\right).
\end{equation}
A random number $x$, drawn from a uniform distribution, is compared to $P_{\rm ul}$; if $P_{\rm ul}>x$, $m_{i, b}$ is considered an upper-limit.

\begin{figure}
    \centering
    \includegraphics[width=\columnwidth]{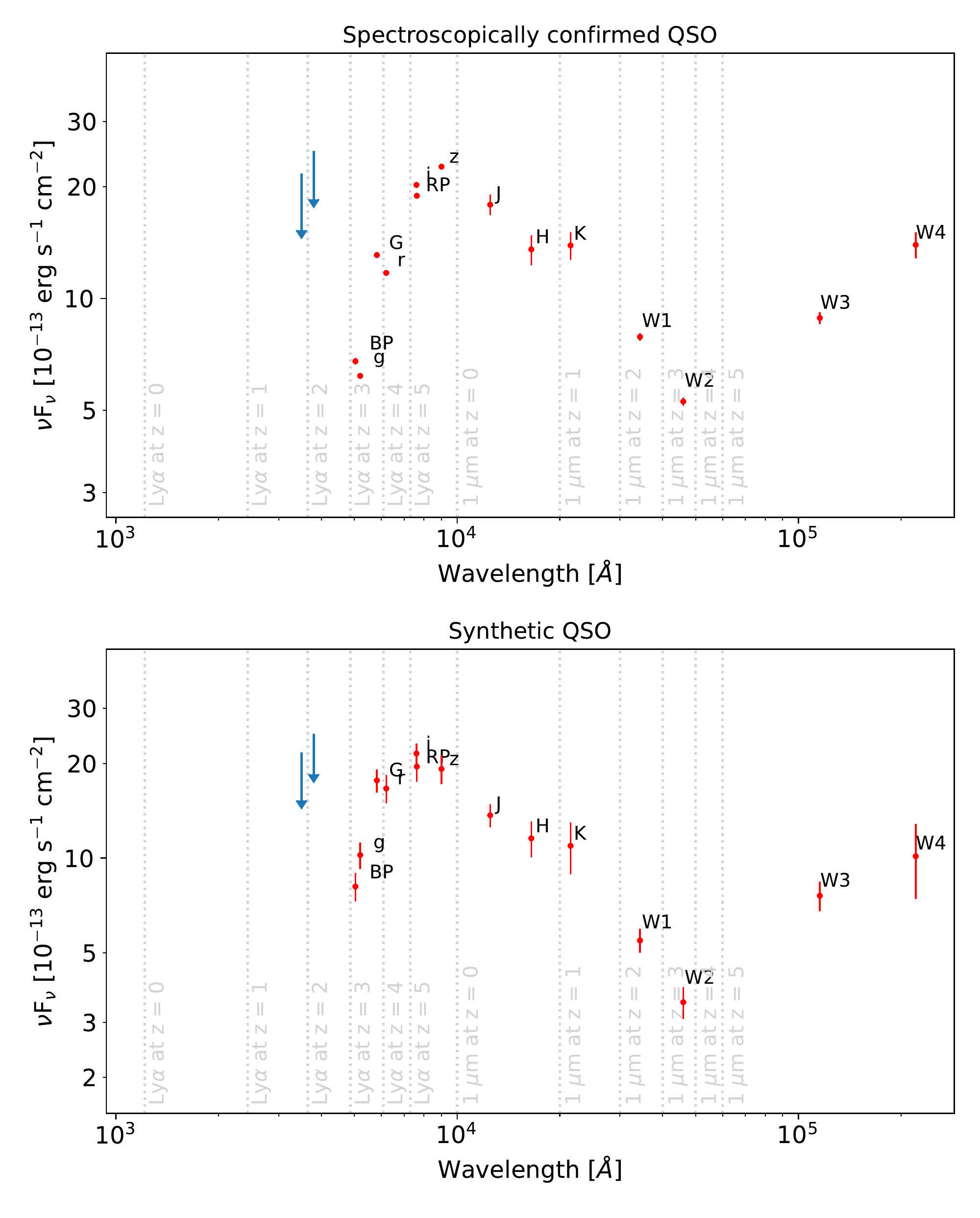}
    \caption{Top panel: SED of one of the spectroscopically confirmed QSOs used as part of the training sample. Bottom panel: a synthetic QSO SED generated with the approach described in Sect. \ref{subsec:SynteticGeneration};  Blue arrows refer to upper limits in the SkyMapper $uv$ bands. Both QSOs are at $z_{\rm em}$ = 3.68.}
    \label{fig:SEDComparison}
\end{figure}

Once upper limits are introduced, the catalogue is ready to be fed to the PRF. The optimal number of synthetic data to employ and their redshift distribution will be discussed in section \ref{sec:testPerformance}. An example of SED of a synthetic object is shown in figure \ref{fig:SEDComparison}; the upper panel shows the SED of a spectroscopically confirmed QSO of comparable redshift and $i_\textrm{psf}$ magnitude.

\section{Performance with synthetic data} \label{sec:testPerformance}
In order to characterise the effect of synthetic data on the selection several tests have been performed. The selection of high-redshift QSOs proceeds in two steps: in the first the PRF is used to differentiate QSOs (at all redshifts) from stars, galaxies and other contaminants. In the second task, instead, we aim to exclude $z<2.5$ targets to obtain a sample of high-redshift QSOs ready for spectroscopic follow-up.

Unless stated otherwise, the same setup was considered: 
i) the available, spectroscopically confirmed high-z QSOs (hereafter also referred to as "real" QSOs) were split in training (75\%) and testing dataset (25\% of the total); the same fractions (75/25) were used for low-z QSOs; additionally, 100 different realisations were drawn for each testing in order to not depend on the particular training-testing choice;
ii) synthetic QSOs were added only to the training dataset, in addition to the spectroscopically confirmed QSOs. The optimal ratio synthetic/spectroscopically confirmed QSOs is not known $a\ priori$: several values were tested in order to identify the configuration that provides the best performance.
This sample was used to understand the capabilities of the PRF in classifying QSOs as low- or high-z candidates.

If training the PRF to classify targets as star, galaxy or QSO at any redshift, stars and galaxies need to be added to the training and testing dataset on top of the QSOs. $Bona fide$ stars were down-sampled based on their $i-z$ colours, i.e., a subset was randomly chosen among all available stars. The number of stars to consider is set by the number of QSOs ($\sim 6000$), in order to maintain a balanced dataset (see e.g., \citet{Paper:UnbalancedJustification:journals/ida/JapkowiczS02} for a review on the class unbalance problem). All spectroscopically confirmed galaxies are instead considered. Stars and galaxies are split according to the 75/25\% rule, and added to the appropriate QSO dataset - training or testing. 

Performance is tested solely on spectroscopically confirmed QSOs (and, when appropriate, known stars and galaxies - see section \ref{sec:QSOStarGal}): synthetic data are never considered in testing datasets when reporting performance values.

The algorithm was evaluated in terms of precision and recall, defined as:
\begin{itemize}
    \item precision: the fraction of relevant instances among all those classified as high-z QSOs. The precision describes how good the algorithm is at excluding contaminants: in the context of this work it describes how good the PRF is at classifying stars, galaxies and low-redshift QSOs as such and not confuse these object as high-redshift QSOs;
    \item recall: the fraction of relevant instance (i.e., real high-$z$ QSOs) that were correctly classified by the algorithm. The recall describes how good the algorithm is at retrieving relevant instances. In the context of this work, it describes how good the PRF is at identifying all high-redshift QSOs present in the MS. 
\end{itemize}

\begin{figure}
    \centering
    \includegraphics[width=\columnwidth]{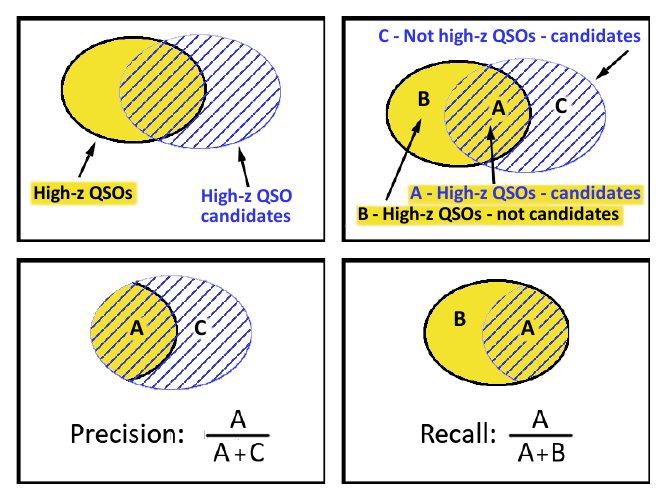}
    \caption{Schematic representation of precision and recall.}
    \label{fig:PrecisionRecallSchema}
\end{figure}

Objects are classified as low- or high- redshift based on a threshold set at $z = 2.5$. However, only when evaluating the recall, two different redshift thresholds, 2.5 and 3.0, are considered. The inclusion of the second higher redshift threshold is justified since targets at $z>3.0$ are the most interesting for our purposes. Moreover, we expect that most of the high-$z$ misclassification to be in the $2.5<z<3.0$ range, as they are closer to the classification threshold. A "global" recall score would fail to highlight the performances on the most interesting objects (those with $z > 3$). 

\subsection{Colour selection}
The selection of high-redshift quasar is often performed by employing colour cuts: optical and infrared colour criteria have been successfully employed in the past for the selection of high-redshfit quasars \citep[e.g.,][]{Onken21:2021arXiv210512215O, Wolf20QSO:2020MNRAS.491.1970W}, and machine learning methods have been successfully employing colours as features \citep[e.g.,][]{Nakoneczny21:2021A&A...649A..81N, QuasarSelection1:2021AJ....162...72W}. Low- and high- redshift QSOs, non active galaxies and stars occupy (mostly) distinct regions in a multidimensional colour space\footnote{There is however some overlapping between star and QSO colours in some redshift intervals, see for instance \citet{SDSSQSOvsStellarColors}.}, allowing an algorithm which employs colours to successfully differentiate between classes. Moreover, we expect the colour distributions to be independent from luminosity, thus allowing to effectively retrieve bright targets that may be under-represented in a training sample.
With respect to Paper I, where only magnitudes were employed as features, in this work we tested colours as well.

\subsubsection{Colour determination}
Given a complete set of magnitudes, colours are trivial to generate, but, due to the introduction of photometric upper-limits, some care is required. In this work all colours are computed as difference between the $b_{\rm th}$ band and the SkyMapper $i$ band, used as a reference. This choice is natural, given that the $i$ band is required for all sources in the main sample. We distinguish three cases:
\begin{itemize}
    \item if the magnitude in the $b_{\rm th}$ band is not an upper-limit, the colour is simply computed as the difference between the measured magnitude and the corresponding $i$ band magnitude. The associated error is the square root of the sum of the photometric errors squared;
    \item if the magnitude in the $b_{\rm th}$ band is instead an upper-limit, we first compute a new $m_{t, n}$ factor (equation \ref{eq:uldist}), specific for each source and photometric band, as the difference between the previously determined value of $m_{t, n}$ and the corresponding $i$ magnitude. The associated error is $\sqrt{\sigma_i^2+\sigma_{t,n}^2}$, where $\sigma_i$ is the error associated with the $i$ magnitude.
    \item if, finally, the magnitude in the $b_{\rm th}$ band is a proper missing data (i.e., it was not observed by a given survey), the colour is marked as a proper missing data as well.
\end{itemize}
Following these three prescriptions, colours are computed both for synthetic and spectroscopically confirmed QSOs.

\subsection{Performance evaluation}
\subsubsection{QSO, Stars and galaxies}
\label{sec:QSOStarGal}
The first task for the selection consists in separating QSOs at any redshift from stars and galaxies. A pre-selection against stars is performed by applying cuts on measured proper motion and parallaxes (i.e., objects with parallax and proper motion significantly different from zero ($>3\sigma$) are considered stars), and against non-active low-$z$ galaxies by excluding extended sources (section \ref{sec:NewMS}).

The same procedure applied in paper I, and briefly summarised in the previous sections, was repeated. The training set is composed of stars, galaxies, low-$z$ and high-$z$ real QSOs and a varying number of synthetic QSOs in roughly equal parts. Stars were randomly under-sampled in order to uniformly cover the entire $i-z$ colour space, while all available QSOs and galaxies were considered. We test the performance of the trained algorithm by applying it on a testing dataset (stars, non-active galaxies and QSOs that were not used during the training) and evaluating the recall and the precision.

The effect of synthetic data was tested, finding no significant improvements in recall and precision over the approach adopted in Paper I. Synthetic QSOs are introduced with $z > z_{\rm th}$ and in varying number. Several redshift threshold $z_{\rm th}$ were tried, from $z_{\rm th} = 1.5$ to $z_{\rm th} = 3.0$ with $\Delta z = 0.25$, as well as different numbers of QSOs, ranging from $N_{synth} = 3000$ to $N_{synth} = 21000$, with $\Delta N_{synth} = 2000$. In all cases, using precision and recall as metrics, the performance measured on the testing dataset are compatible with one another, and with those reported in Paper I. Although non relevant here, synthetic data will become precious to distinguish low-z from high-z QSOs (see section \ref{sec:optimalHyperparmsMags}).
To avoid biases related to a particular testing set, 100 runs were performed, each with a different, random, training/testing split. On average, the PRF achieves a recall of 95\% for QSOs at all redshifts (99\% for QSOs at $z>2.5$, missing 0 or 1 QSO per test run). We measure a precision of 96\%: most of the contaminants are non-active galaxies, accounting for $\sim75\%$ of all misclassifications.

The reported results do not change if training the PRF on colours or on magnitudes data: in this context the PRF performs equally well.

\subsubsection{Distinguishing low from high-redshift QSOs}

\label{sec:optimalHyperparmsMags}
The second step of the selection aims at selecting high-z QSOs: we once again introduce synthetic data in the training dataset in order to improve the generalisation abilities of the PRF at high-redshift.
As reported in Sect. \ref{sec:QSOStarGal}, the performance of the PRF in distinguishing among QSOs, stars and galaxies was comparable either if the algorithm was trained only on spectroscopically confirmed QSOs or on a dataset also including synthetic data. It is thus worth investigating if training the algorithm using the combined dataset provides improvements in terms of predictive capabilities of the PRF for classifying low- and high-redshift QSOs. To address this question, we considered a training dataset composed of both spectroscopically confirmed QSOs and synthetic data in equal parts. All available real QSOs at $z>0.3$ were considered; objects with $0 < z \leq 0.3$ were instead under-sampled to produce a more uniform redshift distribution. The down-sampling strategy was simple: low-$z$ sources were chosen at random among all available targets. After down-sampling, $\sim350$ QSOs with $0 \leq z \leq 0.3$ were kept. 

The PRF was then trained on the combined dataset (i.e., a combination of synthetic and spectroscopically confirmed QSOs) or, alternatively, on the subset containing only spectroscopically confirmed or synthetic QSOs. The predictive performance, measured on a testing dataset of only spectroscopically confirmed QSOs, were compared one with the other. Results are shown in Table \ref{tab:PRF_results_remove_real_or_synth}.
\begin{table*}
    \centering
    \begin{tabular}{l|c|c|c}
    \toprule
    & Recall ($z > 2.5$) & Recall ($z > 3.0$) & Precision \\
    \midrule
    Synthetic and real QSOs & 87.0 $\pm$ 2.5\% & 98.5 $\pm$ 1.0\% & 64.0 $\pm$ 2.0\% \\
    Only synthetic QSOs & 69.5 $\pm$ 3.5\% & 93.0 $\pm$ 2.5\% & 62.0 $\pm$ 3.0\% \\
    Only real QSOs & 67.5 $\pm$ 3.5\% & 83.0 $\pm$ 4.5\% & 82.5 $\pm$ 3.0\% \\
    \bottomrule
    \end{tabular}
    \caption{Recall and precision for a training test with both real and synthetic QSOs, only synthetic or real QSOs. 6000 synthetic QSOs, introduced for $z > 2.5$, were used for the test. The training is always performed by distinguishing low and high-z QSOs, separated at $z = 2.5$.}
    \label{tab:PRF_results_remove_real_or_synth}
\end{table*}
It is clear that a sample of both real and synthetic QSOs produces, in term of recall, the best possible result; improvements in the recall negatively impact the purity of the sample: nonetheless, since the main interest is to improve the completeness of the final sample, both synthetic and spectroscopically confirmed QSOs will be used in section \ref{sec:candidateSample}. We consider the trade-off acceptable, since the number of candidates is still relatively low ($\lesssim 1000$): a different approach might be needed when dealing with a larger number of targets.

\begin{figure}
    \centering
    \includegraphics[width=\columnwidth]{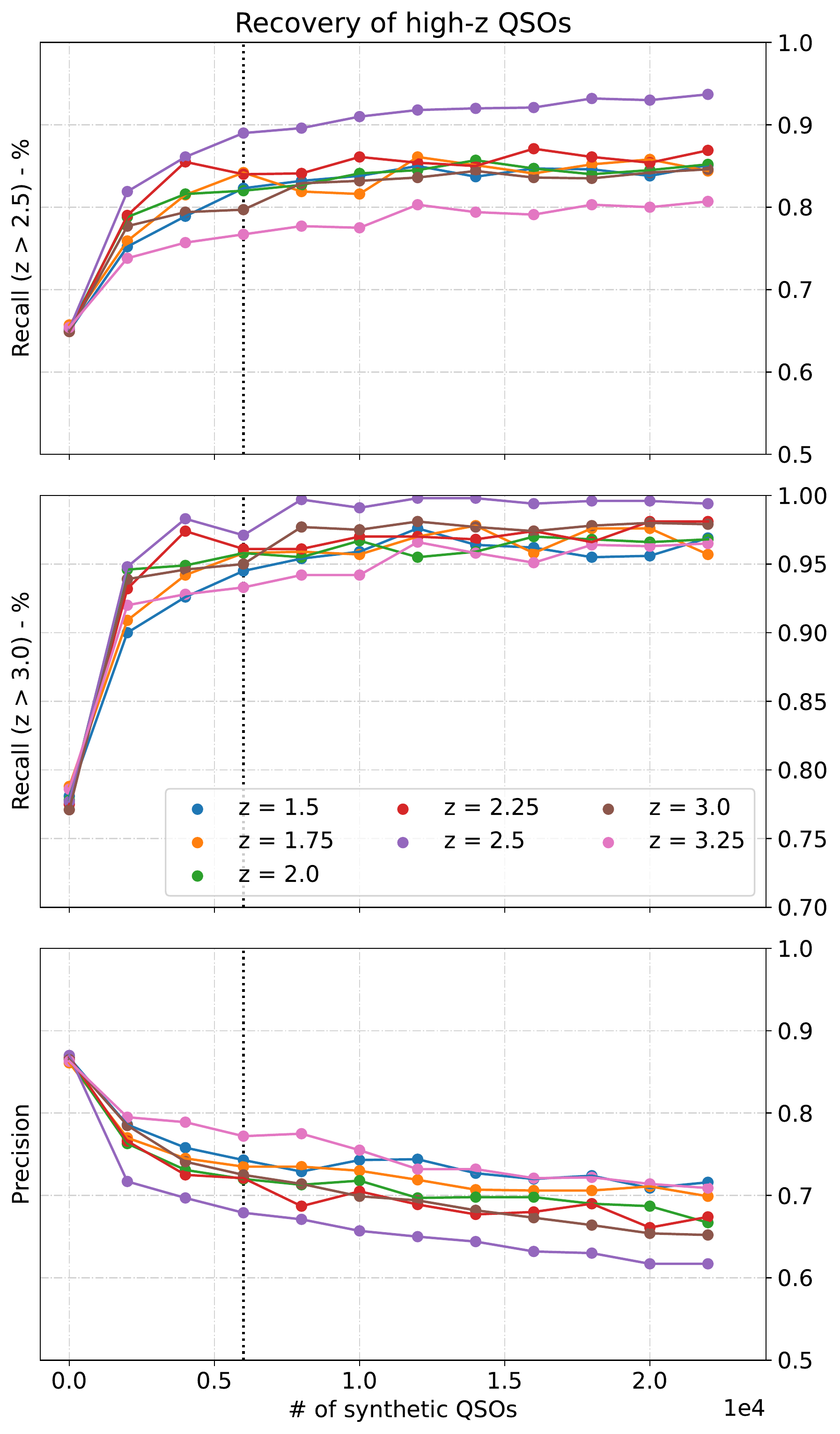}
    \caption{Performance of the PRF in predicting the low-$z$ vs. high-$z$ classification (see text in Sect. \ref{sec:optimalHyperparmsMags}) trained on a dataset with a different number of synthetic QSOs, $N_{synth}$, shown on the x-axis, and different thresholds ($z_{\rm th}$, shown with different coloured lines). The black, dotted line is placed at $N_{synth} = 6000$: in terms of recall at $z>3.0$, no significant improvements are found for $N_{synth}>6000$. We only show the results for the PRF trained on colour data: training the algorithm with magnitudes leads to a slight decrease of precision (from $\sim$2\% to 4\% depending on the chosen parameters), but an overall similar plot.}
    \label{fig:VarySynthParams}
\end{figure}

Once the usefulness of synthetic data is established, the optimal number of generated QSOs to employ and the redshift threshold $z_{\rm th}$ need to be identified. In order to do so, multiple runs of the PRF were considered; for each of these, the number $n_{synth}$ and redshift threshold over which synthetic data are introduced were varied, while keeping the same testing dataset. Results for each pair are illustrated in Figure \ref{fig:VarySynthParams}.

Improvements in the recall are negligible when introducing more than $\sim6000$ synthetic QSOs in the training dataset, independently from the chosen redshift threshold. As a matter of fact, employing more than $\sim6000$ synthetic QSOs degrades the predictive abilities of the PRF in terms of precision. With respect to the number of real QSOs available in the training dataset, $\sim6000$ synthetic QSOs correspond to twice the number of real QSOs at all redshifts: in the following this is the proportion employed for choosing the number of synthetic data to add to the training dataset.  
This result is coherent with the premises of this work: introducing too many synthetic QSOs, mainly at $z>2.5$, produces a training dataset skewed toward high-redshift. As such, the performance of the algorithm will degrade when classifying low redshfit targets, thus lowering the overall precision.

We tested the performance of the PRF using both magnitudes and colours in the training and testing datasets. We found comparable results: the main difference is, on average, a slightly higher precision measured when using colours ($\sim2\%$) with respect to the selection performed with magnitudes. Given the higher purity, the preferred candidate list will be generated with the PRF trained on colour data.

\subsubsection{Leave-one-out test} \label{sec:loo}
Recall and precision are useful to estimate the global performance of the PRF over the entire dataset; due to the low number of very high-redshift QSOs in the dataset they may however paint a biased picture, underestimating misclassification of these targets. To account for this effect a leave-one-out test was performed.
A leave-one-out test consists in training an algorithm on all available data but one object. The latter, left out from the training set, is considered as test. The process is then repeated iteratively for all available objects.
To reduce the computation time needed we restrict the QSOs used for the leave-one-out to the 320 available targets at $z>3.0$. The PRF was thus trained 320 times; in each of these 320 times, the training dataset consists of synthetic data on top of all real QSOs, except one; the testing dataset is just the one of the 320 QSO left out. This procedure allows to test the PRF on all QSOs at $z>3$, one by one. The exercise was repeated with the oversampling approach employed in Paper I as well (see Paper I, end of sec. \ref{sec:catalogue_selectionAlg})

\begin{figure}
    \centering
    \includegraphics[width=\columnwidth]{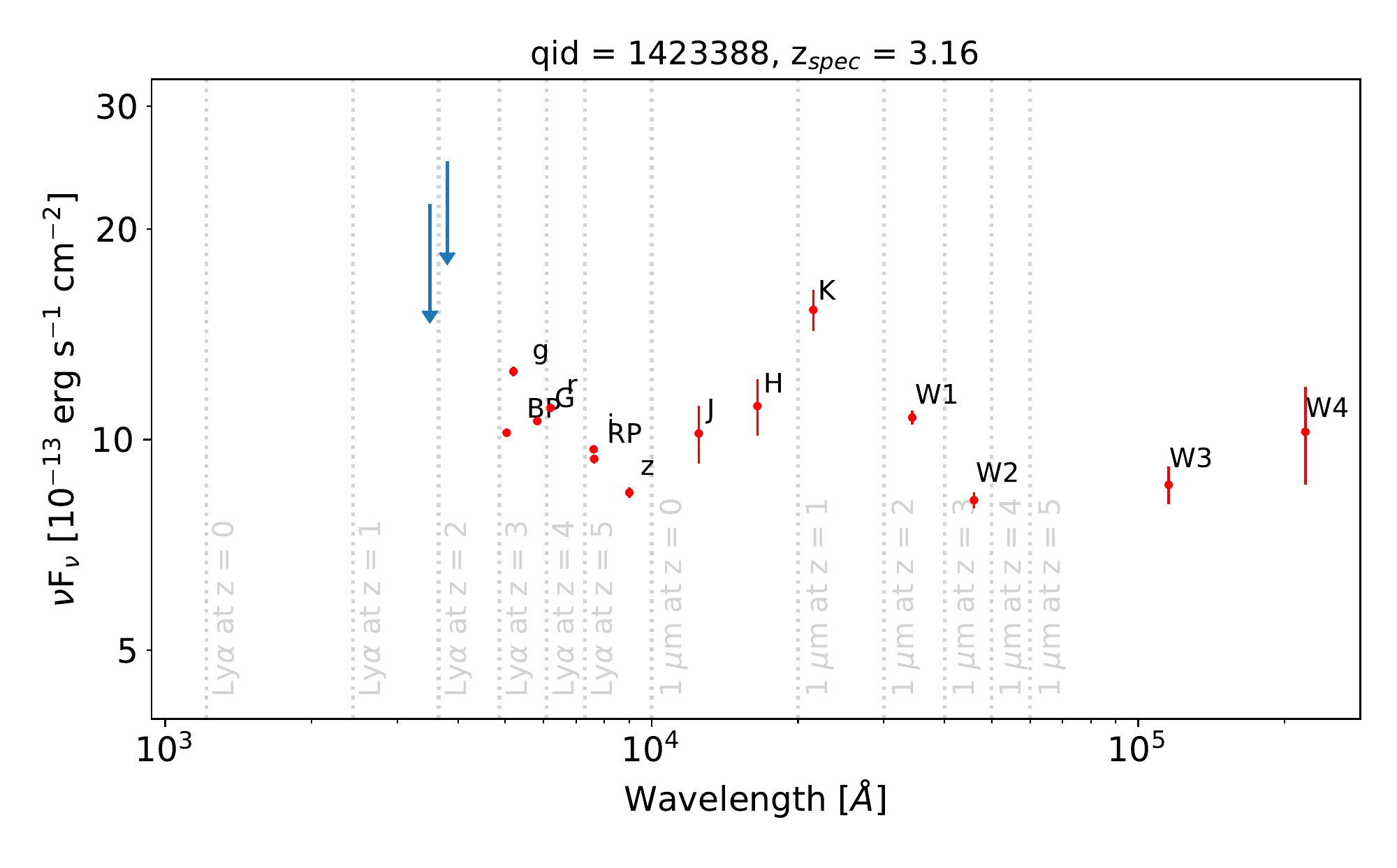}
    \caption{Spectral energy distribution of one of the three objects misclassified (QID = 1423388, $z_{\rm em} = 3.16$) by the PRF when trained both with colours and magnitudes. Blue, downward arrows denote upper limits in the SkyMapper $u$ and $v$ filters.}
    \label{fig:peculiarSED}
\end{figure}

Out of the 320 available QSOs at $z>3$, we find that the PRF misclassifies 4 targets when trained on synthetic data, 13 with the method described in Paper I. Considering the results at face value, both approaches work rather well, with at most 13 ($\sim4\%$ of the total) misclassifications. A more careful analysis however shows that the algorithm trained with synthetic data produces better results:
\begin{itemize}
    \item the number of misclassified targets is lower (4, compared to the 13 misclassified registered with the method outlined in Paper I). 3 of the misclassifications are in common: two of them have peculiar SED, which are not similar to the typical SED of high-$z$ QSOs (Fig. \ref{fig:peculiarSED}). IR spectra would be required to understand whether this peculiarity is intrinsic or, e.g., due to effect of variability. The SED of the third object, on the other hand, is rather typical; the most likely explanation for its misclassification as low-$z$ QSO is a detection in the SkyMapper $v$ band, extremely uncommon in the sample for quasars with $z\sim3$, possibly an indication that the line of sight toward this target is particularly unabsorbed.
    
    \item the method outlined in paper I misclassifies 4 out of the 13 available targets at $z>4.5$, which are instead recovered by the PRF trained on synthetic data;
    \item a bright, high-$z$ target, recently identified with a novel selection approach (Calderone et al. in preparation) is not selected by the method outlined in paper I, while it is correctly classified when training the PRF with synthetic data. Misclassifying this target is especially undesirable, as it is both bright and at high-redshift: it is thus an ideal candidate for the redshift drift Sandage test \citep[e.g.,][]{Liske+08:2008MNRAS.386.1192L, Boutsia20:2020ApJS..250...26B}, the search of variation for fundamental constants \citep[e.g.,][]{FundConstMurphy:ApJ, FundConstMilakovic:2021MNRAS.500....1M}, or any other high resolution, precision and stability experiment to be carried out in the future.
\end{itemize}

The Leave-One-Out test was repeated with the PRF trained both on colours and on magnitudes, finding the same results reported above.

\subsection{Literature misclassifications} \label{sec:literatureMiscs}
\begin{figure}
    \centering
    \includegraphics[width=\columnwidth]{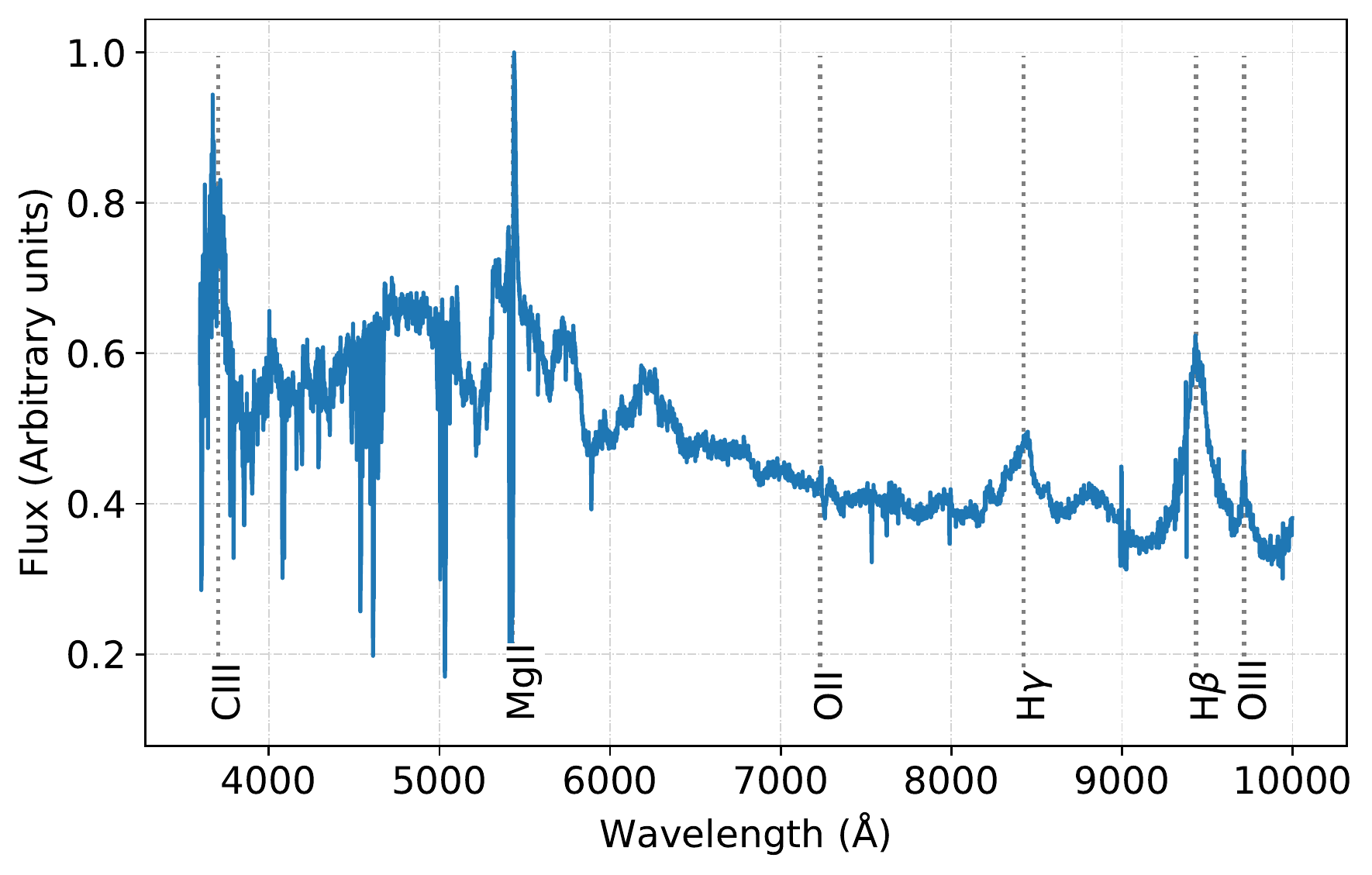}
    \caption{Example of an object from the SDSS with an incorrect redshift estimate. The $H\beta, \textrm{MgII}, \textrm{H}\gamma, \textrm{OIII}$ lines are visible, constraining the redshift at $z = 0.94$ with respect to the SDSS quoted redshift of 3.457.}
    \label{fig:literatureMisclassificationsExample}
\end{figure}

As a byproduct of the tests described in the previous sections we identified some misclassifications in published catalogues. We report them in table \ref{tab:literatureMisclassifications}, together with a revised redshift and/or class. An example of a spectrum for a misclassified target is shown in Fig. \ref{fig:literatureMisclassificationsExample}.

These targets were identified by visual inspection of the misclassified objects in a given testing/validation dataset. After each evaluation run, the spectral energy distribution and, if available, the spectrum of misclassified targets at $z>2.5$ is visually inspected. This is done to try to understand the reason for the misclassification and to check whether the target is genuinely a high-redshift QSO or was assigned an incorrect redshift estimate or class in the original catalogue. Following this approach, 4 targets with an available spectrum were re-classified or given a new redshift estimate.

Additionally, outliers identified as objects with a high reduced $\chi^2$ of the synthetic photometry with respect to the observed magnitudes while building the composite SED (sect. \ref{sec:meanSED}) were visually inspected, resulting in the discovery of 5 additional errors, in the redshift determination and/or in the object classification; they are also reported in table \ref{tab:literatureMisclassifications}.

\begin{table*}
    \centering
    \begin{tabular}{c|c|c|c|c|c|c|c|c}
    \toprule
         QID & RA (J2000) & DEC (J2000) & $z_\mathrm{new}$ & $z_\mathrm{literature}$ & Obj. type & Obj. type & Original catalogue & Method \\
             &            &             &                  &                         & Literature & Revised  & & \\
         \midrule
         1121192 & 01:40:04.44 & -15:32:55.68 & 0.819 & 1.669 & QSO & QSO & Veron10    & PRF \\
         1351340 & 21:10:55.92 &  05:07:07.97 & 0.94  & 3.457 & QSO & QSO & SDSS DR16Q & PRF\\
         1353218 & 00:27:49.94 &  07:06:40.25 & 0.84  & 3.201 & QSO & QSO & SDSS DR16Q & PRF\\
         1362635 & 16:00:26.16 &  00:28:34.17 & 0.124 & 3.76  & QSO & Galaxy & Veron10 & PRF\\
         2223596* & 14:48:25.80 &  10:31:57.83 & 0.0  & 5.864 & QSO & Brown dwarf & Veron10 & SED \\
         2223725 & 14:53:31.17 &  14:21:12.68 & 0.07  & 1.072 & QSO & Galaxy & Veron10 & SED\\
         2223946 & 15:02:58.01 &  13:18:52.93 & 0.0   & 0.659 & QSO & Galaxy & Veron10 & SED\\
         2224363* & 15:38:43.10 &  08:42:37.00 & 0.0   & 2.735 & QSO & White Dwarf & Veron10 & SED\\
         \bottomrule
    \end{tabular}
    \caption{Misclassification identified while running the PRF on testing datasets or visually inspecting outliers with respect to the average quasar SED. The "Method" column clarifies the method which flagged each target for review.\\ * the target was also noted as a misclassification in Simbad.}
    \label{tab:literatureMisclassifications}
\end{table*}

\section{The candidate samples}
\label{sec:candidateSample}
\begin{figure}
    \centering
    \includegraphics[width=\columnwidth]{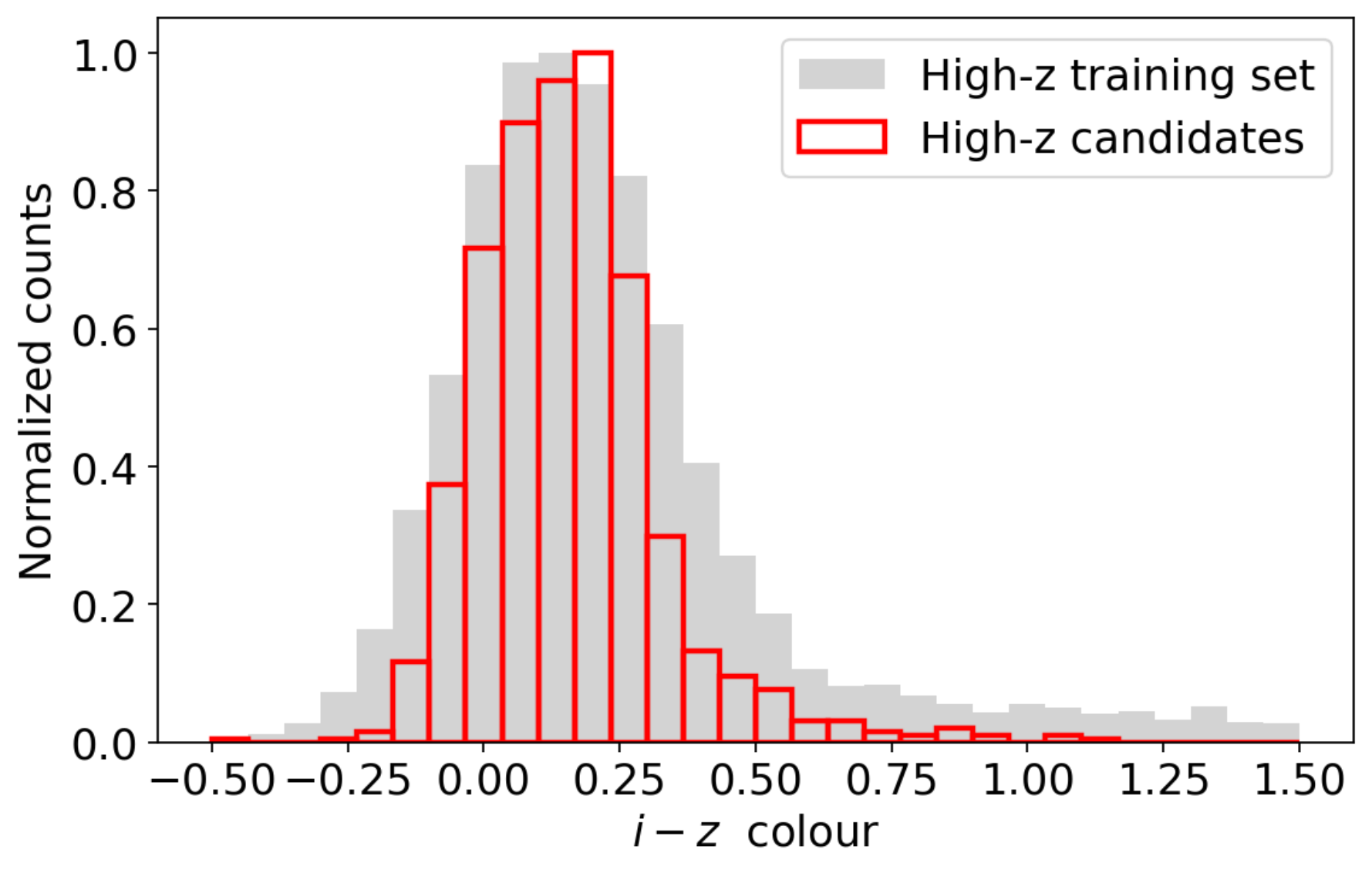}
    \caption{$i-z$ colour distribution for QSO with $z>2.5$. The grey histogram in the background shows the distribution for the training set, while the red histogram the distribution for high-$z$ candidates; both histograms have been normalised to unity, i.e., the most populated bin was normalised to unity.}
    \label{fig:magnitudeDist}
\end{figure}
The previous sections described the tests performed on known data in order to understand how well the algorithm performs in the selection of high-redshift QSOs. We now turn to the application of the algorithm on unlabelled data, in order to extract a new candidate list.

Two lists were produced: one obtained from a training sample where magnitudes (both of real and synthetic QSOs) are used as features; another where, instead, colours are employed. Based on the results presented in the previous sections, the two lists are expected to be similar.

After removing likely extended targets \citep[][]{Calderone19:2019ApJ...887..268C}, \nNonClassfied{} out of the \totalMSsources{} sources in the original main sample are still lacking a classification label. As a first step, the PRF is trained to separate QSOs, stars and non-active galaxies (Sec. \ref{sec:QSOStarGal}). When applied on the dataset of unclassified sources, the algorithm produces a list of \nQSOCandidates{} (\nQSOCandidatesColors{}) QSO candidates at all redshifts if trained on magnitudes (colours), \nStarCandidates{} (\nStarCandidatesColors{}) star candidates and \nGalCandidates{} (\nGalCandidatesColors{}) non-active galaxy candidates. The lists of stars and non-active galaxies are discarded: they are, for our purposes, contaminants. During the second step, and given the sample of QSO candidates at all redshfit, the algorithm is trained on QSO data in order to distinguish low- and high-redshift targets (sec. \ref{sec:optimalHyperparmsMags}). The list consists of \candMag{} candidates if trained with magnitudes, \candCol{} when the algorithm is trained with colours; the two lists share $\sim$85\% of the targets. 

Both selections were performed using the optimal hyper-parameters identified in section \ref{sec:testPerformance}. Consistently with the results discussed in previous sections we find similar completeness ($\sim87\%$) for both lists, but slightly lower contamination from the candidate list generated by training the PRF on colour data. This is consistent with the number of candidates recovered, assuming that a large part of the non-overlapping candidates are contaminants (either galaxies, stars or low redshift targets).

As a consistency check, we compared the colour distribution of the high-redshift QSO candidates with that of the training sample. We expect the distribution of the training set, which includes synthetic data, to extend to larger values of a given colour: synthetic data were introduced to provide a more extensive coverage in the colour space. The $i-z$ colour was chosen, as all sources in the main sample are required to have a $i$ magnitude measurement, and all but 32 targets, mostly stars, have a $z$ magnitude measurement. Figure \ref{fig:magnitudeDist} compares the distribution of colours for the training set and the high-$z$ candidate list, obtained when training the PRF on magnitudes: as expected, the two are consistent with each other but with a wider distribution for the training set. Similar results are found when training the PRF on colours, as most of the candidates are in common between the two lists.

\section{Spectroscopic validation and catalogue} \label{sec:observations}
New spectra of quasar candidates are continually taken in the framework
of the QUBRICS program.
To maximise the success rate and reduce the contamination, targets are selected following an "iterative" approach:
after each spectroscopic run the training sample is updated to include the new identifications.
In this way the training set and the list of candidates is evolving after each run:
for the purposes of this paper they have been frozen at the situation at the end of February 2022.

In Table \ref{tab:NewSpec}  we list the new spectroscopic identifications - not reported in the previous QUBRICS  papers -
of PRF candidates from the list of Paper I and updated lists based on the SkyMapper DR3. 
They are part of the training sample described in this work.

\begin{figure*}
    \centering
    \includegraphics[width=.9\textwidth]{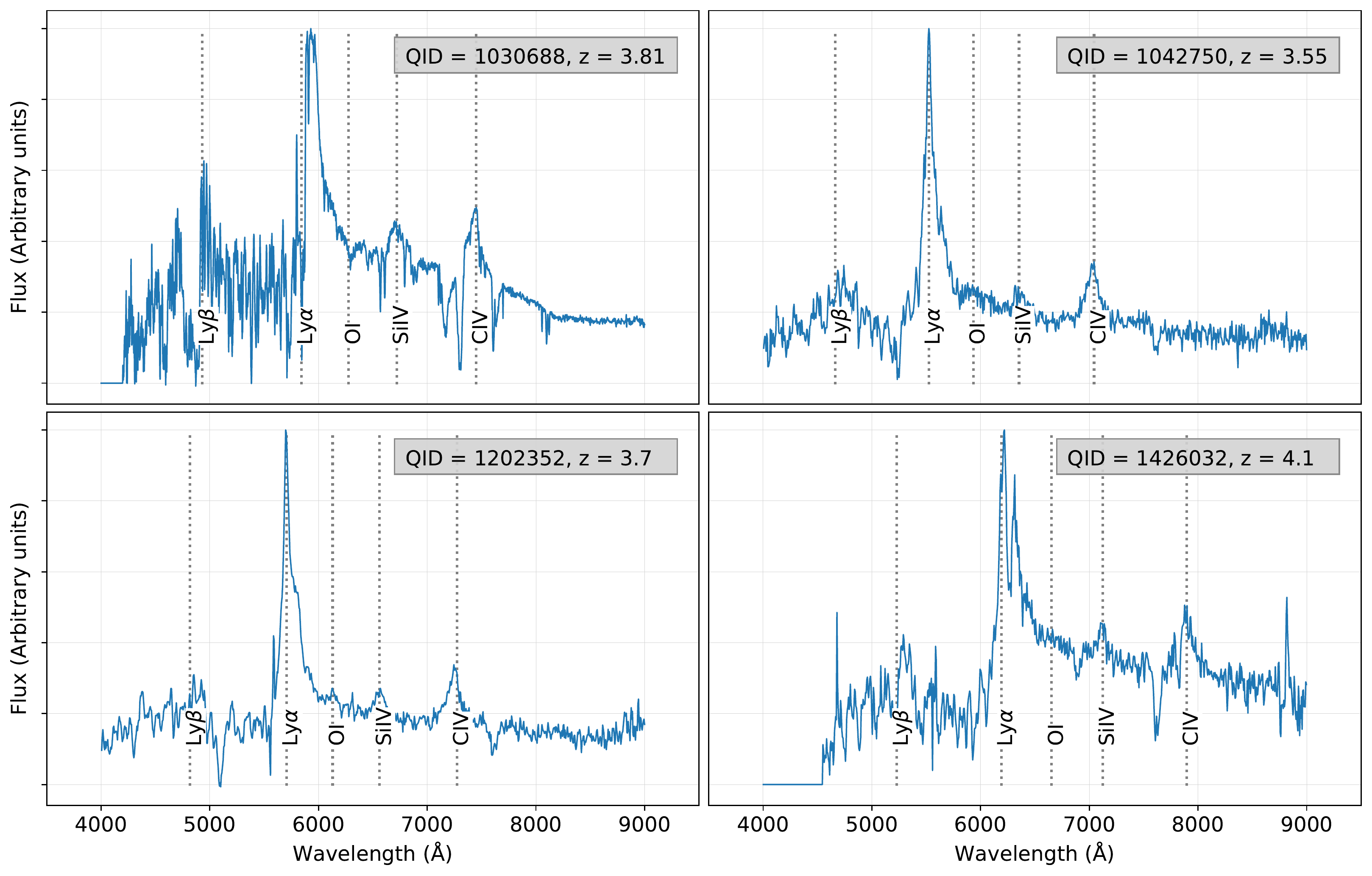}
    \caption{Example of discovery spectra for 4 of the QSO in the published catalogue. Spectra were taken with the EFOSC2 and LDSS3 spectrographs.}
    \label{fig:exampleSpectra}
\end{figure*}

A total of 206 PRF-selected candidates have been observed at Las Campanas Observatory, Telescopio Nazionale Galileo (TNG, La Palma) and ESO-NTT using respectively the LDSS-3 (Clay Telescope), DOLORES and EFOSC-2 spectrographs. 
Table \ref{tab:observations} summarises the observing setups and significant information about each observing run. Figure \ref{fig:exampleSpectra} shows some selected spectra drawn from the catalogue.

109 of the 206 presented candidates have been observed with LDSS-3 at the Clay Telescope. Observations were obtained in several nights with varied conditions (e.g., bright time, variable weather conditions). The VPH-all grism with the 1" central slit and no blocking filter was employed, covering a wavelength range between 4000--10000 {\AA} with a low resolution of R$\sim$800. Exposure times ranging between 800--1800 s were used, depending on the candidate magnitude. Data obtained from the LDSS-3 instrument were reduced with a custom pipeline based on MIDAS \citep[][]{ref:midas} scripts. Each spectrum has been processed to subtract the bias and normalised by the flat; wavelength calibration is achieved using helium, neon and argon lamps, finding an rms of $\sim0.5${\AA}. Observing conditions have not always been photometric: the flux calibration is thus relative to a spectro-photometric standard star, observed each night.
\begin{table}
    \centering
    \begin{tabular}{c|c|c|c|c|c}
    \toprule
        \# of objects & Instrument & Telescope & Grism & Slit & Resolution \\
        \midrule
        109 & LDSS-3  & Clay   & VPH-all   & 1"   & 800  \\
        80  & DOLORES & TNG    & LR-B      & 1"   & 600  \\
        17  & EFOSC-2 & NTT    & Grism\#13 & 1.5" & 1000 \\
        \bottomrule
    \end{tabular}
    \caption{Setup for each observing run.}
    \label{tab:observations}
\end{table}
Eighty candidates were observed with the DOLORES instrument mounted on the Telescopio Nazionale Galileo. Exposures have been taken during the AOT43 and AOT44 periods, from April to November, under two proposals (PI: F. Guarneri); the LR-B grism (resolution $\sim600$) with a 1" slit aperture was used with an exposure time between 300 and 600 s.

In November 2021 we were awarded four nights (PI. F. Guarneri, proposal 108.22L1.001) at NTT, employing the EFOSC-2 instrument Grism \#13 (wavelength range $\lambda\sim 3700-9300$ {\AA}), with typical exposure times ranging between 400 and 800 s; 17 candidates from the list obtained with Paper I and the present work were observed and given a robust identification and redshift. The same data reduction procedure applied for LDSS-3 data was used for the EFOSC-2 spectra.

Out of the 206 targets, 149 turned out to be genuine QSOs with $z > 2.5$, 41 QSOs with z$ < 2.5$ ($<z> = 2.18$), 3 galaxies and 13 stars. Based on the results of these runs, the achieved success rate is of 72\%, with the most significant contaminant being low-redshift QSOs. Among the 41 low-redshift QSOs, 7 show broad absorption features in their spectra, and have been marked as * in table \ref{tab:NewSpec}; BAL features mimic the colours of high-$z$ QSOs, making it hard to select against these particular targets \citep[][]{ref:cupaniFeLoBALs}.

\section{Estimated success rate of the remaining candidates}
\label{sec:estimatedSuccessRate}
\begin{figure}
    \centering
    \includegraphics[width=\columnwidth]{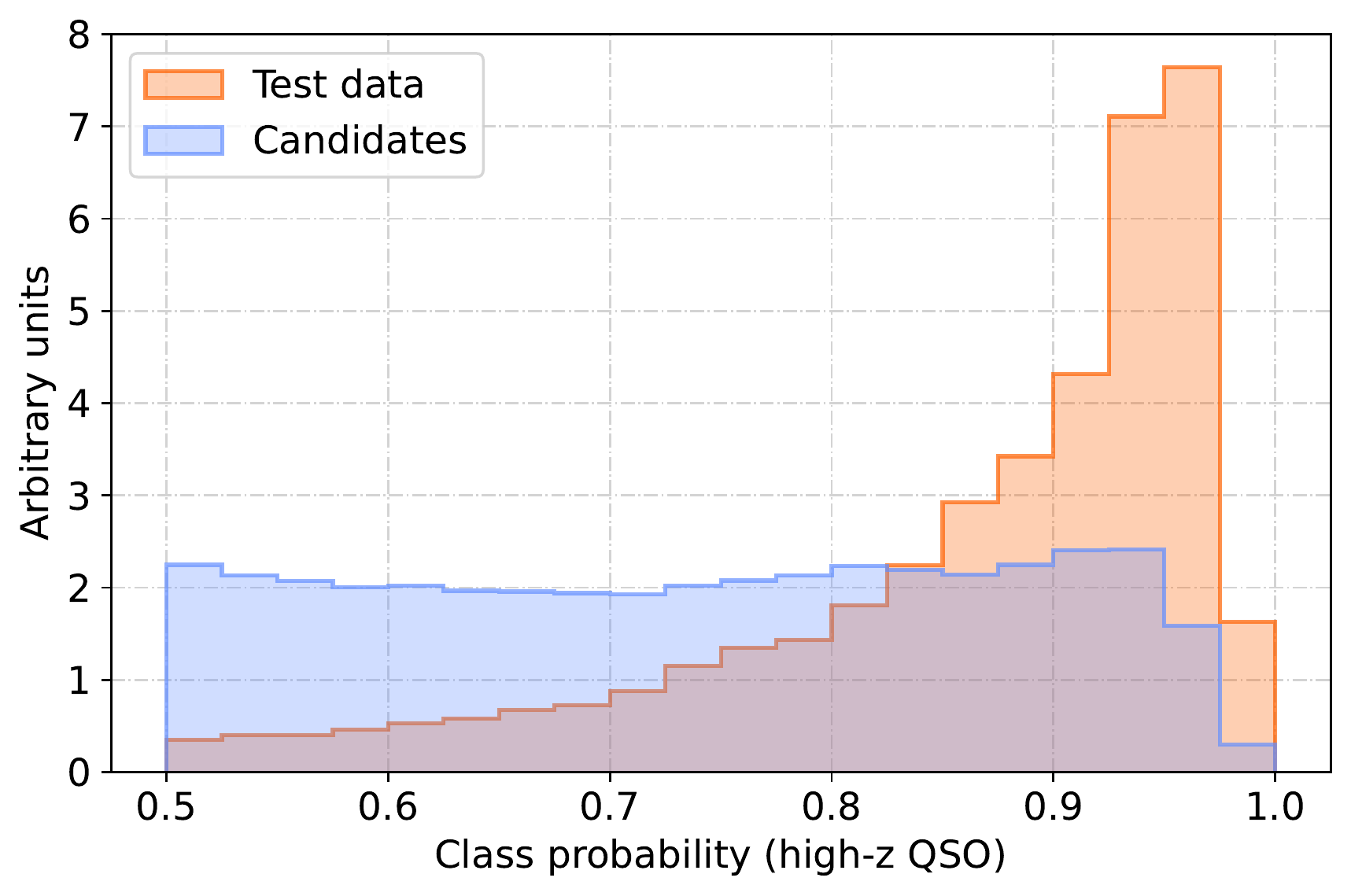}
    \caption{Class probability (high-z QSO) of QSO candidates in the testing dataset and in the candidate list, obtained with the PRF trained on colour data. The light blue histogram shows the probability distribution of high-$z$ QSO candidates, while the orange one the high-z class probability distribution of the testing set. A class-probability lower that 0.5 implies that a source is classified as low-$z$ QSO, and as such is not shown in this plot. Both histograms have been normalised so that the integral over the range is equal to one.}
    \label{fig:ReliabilityComparison}
\end{figure}
The precision of our algorithm for high-$z$ QSOs selection has been estimated in Sect.~\ref{sec:testPerformance}. However, if we were to observe all the remaining candidates, we expect a lower precision since (i) during candidate prioritisation we already observed the most promising ones and (ii) the adoption of synthetic spectra for training may bias our estimate.  In this section we will estimate a more reliable precision for the remaining candidates; this estimate is equivalent to the success rate expected on the candidate sample.

As mentioned in Sect. \ref{sec:catalogue_selectionAlg}, the PRF computes, for each object, both a class label and the probability of belonging to the assigned class. This probability can be interpreted as a reliability score: the higher the probability, the most likely a candidate is an actual high-$z$ QSO. Comparing the probabilities distribution  of high-$z$ QSO candidates with those of a testing dataset (Fig. \ref{fig:ReliabilityComparison})
it is clear from the blue histogram that a significant part of the objects in the candidate list have a lower probability of being high-$z$ QSOs with respect to the testing dataset. In order to keep this into account we use the QSO luminosity function to estimate the expected number of high-$z$ QSOs in the candidate list and estimate the success rate of a telescope run observing all the \candCol{} targets  selected in the previous section with the PRF trained on colour data, 

As in paper I, we assume that the PRF is able to exactly select QSOs, stars and non active galaxies. 

We first try to estimate how many non-QSOs are expected to contaminate the candidate list. The PRF is thus trained to distinguish QSOs, stars and non-active galaxies. Applied on a testing dataset (the same used in section \ref{sec:QSOStarGal}), a small fraction of the stars ($\lesssim1\%$) and galaxies ($\sim$ 6.5\%) into it are misclassified as QSOs.
Following the steps described in the previous sections these sources, actually contaminants, would then be re-processed to identify low and high-redshift targets. As such, the algorithm is trained to distinguish low- and high-$z$ QSOs, and used to label the stars and non-active galaxies just misclassified as QSOs. This leads to $\sim1\%$ of the galaxies and $\lesssim0.1\%$ of the stars initially present in the testing dataset to be labelled as high-redshift QSO. 
We then multiply these estimates with the predicted number of stars and galaxies among the unclassified sources. This should provide the number of stars and galaxies expected to contaminate the candidate list, which turns out to be $\sim10$ galaxies and $\sim20$ stars.

\begin{figure}
    \centering
    \includegraphics[width=\columnwidth]{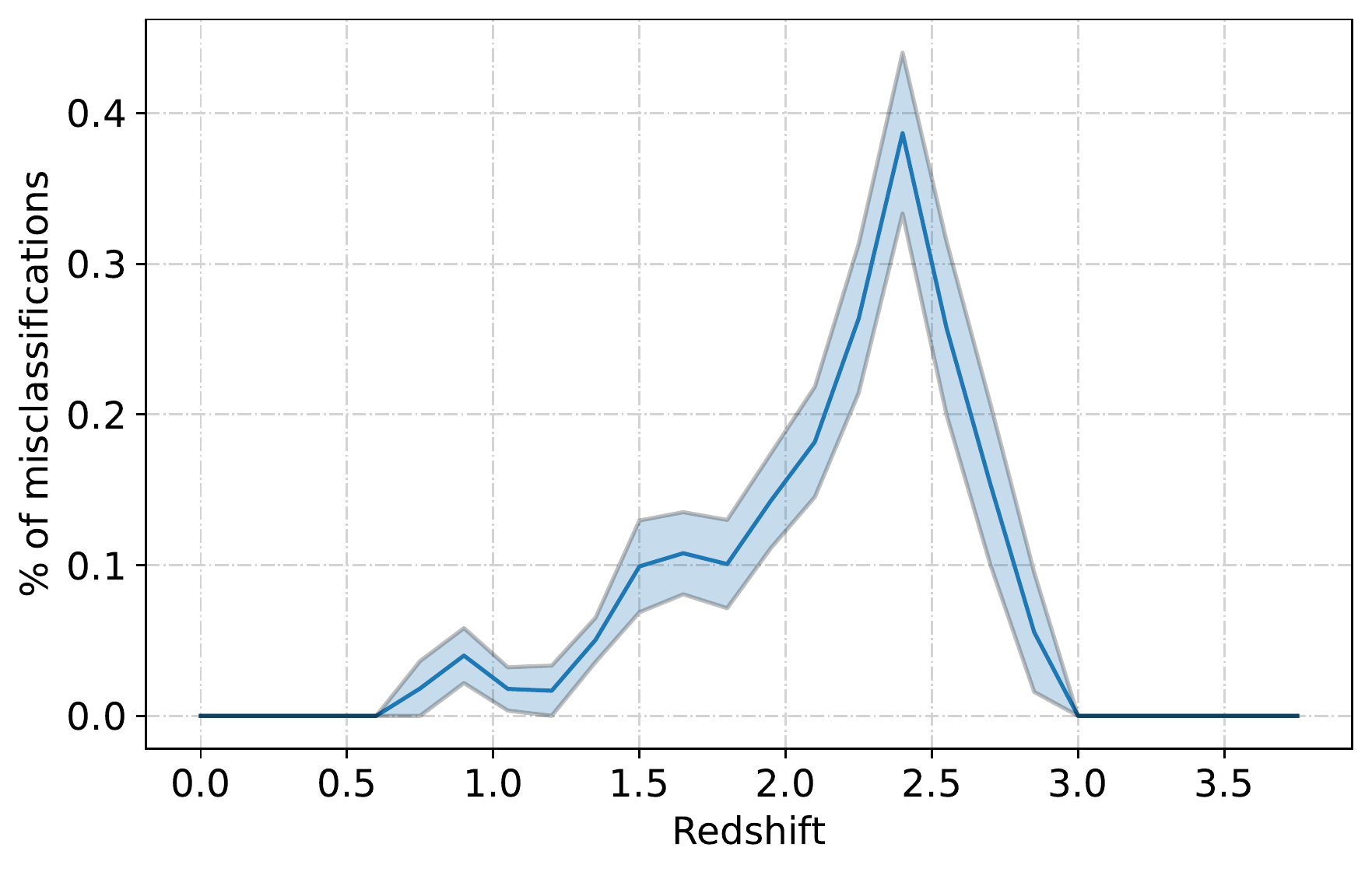}
    \caption{Median misclassifications rate as a function of redshift; the shaded area represents the MAD. Both the median and the MAD are calculated over 100 randomly chosen testing datasets.}
    \label{fig:misclassificationRate}
\end{figure}

The most significant contaminant is however represented by low-redshift QSOs \citep{Calderone19:2019ApJ...887..268C, Boutsia20:2020ApJS..250...26B}: in Paper I, these were estimated to be $\sim85\%$ of all contaminants. 
In order to assess their significance for the current candidate list, we multiply the expected number of QSOs as a function of redshift, derived from \citet{Kulkarni19} and corrected at high-redshift with \citet{LF_Hz_Grazian:2021arXiv211013736G, LF_Boutsia:2021ApJ...912..111B} with the misclassification rate. We define the latter as the ratio between the number of misclassfied QSOs in a given redshift bin and the total number of QSOs in the same redshift bin.
In order to calculate the misclassification rate, we train the PRF to separate low- and high-redshift QSOs; to avoid biases due to a particular choice of training/testing datasets, 100 different iterations were considered, each with a different training/testing split.
We calculate the misclassification rate separately for each of these runs, and a global misclassification rate as the median across all 100 iterations; an uncertainty is estimated as well by using the median absolute deviation. The results are shown in figure \ref{fig:misclassificationRate}. Most of the misclassifications happen at $2\leq z\leq2.5$, close to the threshold separating the two classes.

Keeping into account the misclassification rate shown in Fig. \ref{fig:misclassificationRate} and that, as suggested by \citet[]{LF_Boutsia:2021ApJ...912..111B}, SkyMapper has an inherent incompleteness of $\sim15\%$ with respect to QSO data, we expect to find $\sim590$ low redshift QSOs in the candidate sample, mostly in the $2 < z < 2.5$ redshift range.
In this way the success rate for the candidates still to be observed is expected to be around $31\%$ with about $278$ new QSOs at $z>2.5$ and $i<18$ to be found.

\section{Discussion and conclusion} \label{sec:conclusions}
The probabilistic random forest is an effective tool in selecting bright ($i<18$), and high-redshift ($z>2.5$) QSOs: in \citet{Guarneri:2021MNRAS.506.2471G} we presented a first approach to the selection of these targets based on photometric data.
In this paper we aimed at improving the selection method outlined in Paper I by exploring the effect of synthetic data used as training set.
Synthetic data are commonly used when training machine learning algorithms, and their role is expected to become even more prominent in future years \citep[e.g.,][]{Nikolenko2021}. Diversified training examples can make a ML algorithm more precise,  including corner cases and under-represented samples to make up for the lack of examples found in real-world datasets. Bright, high-$z$ QSOs are a clear example of under-represented targets: training a ML algorithm for identifying them can prove difficult; introducing synthetic data significantly eases the task.

In this work synthetic data have been generated based on the mean properties of previously spectroscopically known QSOs, including new QUBRICS identifications. A composite spectral energy distribution was computed, and used to obtain synthetic magnitude measurements by multiplying the estimated fluxes by instrumental pass-bands. The candidate selection was split in a two-step process: in the first step we aim at sifting QSOs, at all redshifts, from stars and galaxies. In the second, we re-classify QSO candidates in order to distinguish low- and high- redshift targets. Synthetic data are used for training the algorithm during this second step. In all cases, we use precision and recall as evaluation metrics for the algorithm.

\begin{itemize}
    \item The comparison of the performances of the new approach with respect to Paper I shows significant improvements in recovering a large part of the bright and/or very high-redshift targets.
This is especially significant, as bright and $z>3$ QSOs are the most precious targets for future experiments, such as the search for varying fundamental constants \citep[][]{FundConstMilakovic:2021MNRAS.500....1M, FundConstMurphy:ApJ} or the redshfit drift experiment \citep[][]{Liske+08:2008MNRAS.386.1192L, Boutsia20:2020ApJS..250...26B}.
    The improvements in recall cost something in terms of precision, lowering the overall score from 78\% to 64\%. Since we are mostly interested in improving the recall and the number of QSO candidates remains manegeable, we consider this trade-off acceptable;
    \item the effect of colours on the selection was tested, with and without synthetic data. We found slightly better results when colours are adopted, mainly in the precision metric, with respect to those obtained when using magnitudes as features. Similarly to the results reported above, synthetic data significantly improve the measured recall also when colours are adopted.
\end{itemize}

We stress that precision and recall are only used to evaluate the performance of the algorithm and should not be intended as, respectively, success rate of the survey and completeness of the final sample. In particular:
\begin{itemize}
    \item we use precision as the evaluation metric, but note that the expected success rate on the remaining candidates may be lower (Sect.~\ref{sec:estimatedSuccessRate}). The two are not expected to be equal to each other: due to the iterative nature of the survey, as the survey progresses new spectroscopically confirmed QSOs are included in the training set and are thus removed from the candidate list. The success rate measured on the sample will thus drop with time, as more spectroscopically confirmations are carried out; 
    \item we chose the recall, rather than the QSO completeness, as metric to characterize the algorithm since the former depends only on the selection performances, while the latter depends also on the MS selection. More specifically, in order to estimate the completeness we would also needs to keep into account the number of quasars excluded \textit{a priori} from the MS due to the requirements mentioned in section \ref{sec:NewMS}. According to \citet[][]{LF_Boutsia:2021ApJ...912..111B}, the MS has an inherent incompleteness of 15\% with respect to QSO data, we expect the completeness of the selection to be $\sim$70\%, i.e., 0.85 times the recall, estimated to be $\sim85$\%.
\end{itemize}

The selection method outlined in this work was applied on unlabelled data from a new main sample, based on the third data release of the SkyMapper survey. The data include photometric measurements from the SkyMapper Dr3, Gaia eDR3, AllWISE and 2MASS surveys. When trained on colour data, the algorithm produces a list of \candCol{} candidates.

A large number of spectroscopic follow-ups have been carried out since Paper I, and 206 targets have been observed. These 206 have been selected in the past by the PRF, and are now part of the training sample: in order to maximise the success rate, we follow an iterative approach, updating the QSOs candidate list after each observing run.
Of the 206 targets, 149 turned out to be genuine high-$z$ QSOs.
About 278 more QSOs at $z>2.5$ and $i<18$ are expected to be contained in our present
candidate list.

We are further refining the selection method and continuing the spectroscopic campaign, in order to expand the spectroscopic sample and provide more training to our machine learning algorithms. Besides, novel selection techniques are being developed and currently tested, while new photometric datasets are also being employed
(Calderone et al., in preparation).

%
\section*{Acknowledgements}
We thank M. Fumagalli, J. X. Prochanska and J. Hennawi for the SpecDB database \citep{specDB}, and for kindly providing spectra of previously observed targets. We thank an anonymous referee for a timely and thorough report that helped to clarify various aspects of the paper.
We thank Societ\`a Astronomica Italiana (SAIt), Ennio Poretti, Gloria Andreuzzi, Marco Pedani, Vittoria Altomonte and Andrea Cama for the observation support at TNG. Part of the observations discussed in this work are based on observations made with the Italian Telescopio Nazionale Galileo (TNG) operated on the island of La Palma by the Fundacion Galileo Galilei of the INAF (Istituto Nazionale di Astrofisica) at the Spanish Observatorio del Roque de los Muchachos of the Instituto de Astrofisica de Canarias.

The national facility capability for SkyMapper has been funded through ARC LIEF grant LE130100104 from the Australian Research Council, awarded to the University of Sydney, the Australian National University, Swinburne University of Technology, the University of Queensland, the University of Western Australia, the University of Melbourne, Curtin University of Technology, Monash University and the Australian Astronomical Observatory. SkyMapper is owned and operated by the Australian National University’s Research School of Astronomy and Astrophysics. The survey data have been processed and provided by the SkyMapper Team at ANU. The SkyMapper node of the All-Sky Virtual Observatory (ASVO) is hosted at the National Computational Infrastructure (NCI). Development and support the SkyMapper node of the ASVO has been funded in part by Astronomy Australia Limited (AAL) and the Australian Government through the Commonwealth’s Education Investment Fund (EIF) and National Collaborative Research Infrastructure Strategy (NCRIS), particularly the National eResearch Collaboration Toolsand Resources (NeCTAR) and the Australian National Data Service Projects (ANDS).

This work is based on data products from observations made with ESO Telescopes at La Silla Paranal Observatory under ESO programmes ID 108.22L1.

This work has made use of data from the European Space Agency (ESA) mission Gaia (https://www.cosmos.esa.int/gaia), processed by the Gaia Data Processing and Analysis Consortium (DPAC, https://www.cosmos.esa.int/web/gaia/dpac/consortium). Funding for the DPAC has been provided by national institutions, in particular the institutions participating in the Gaia Multilateral Agreement.

This publication makes use of data products from the Two Micron All Sky Survey, which is a joint project of the University of Massachusetts and the Infrared Processing and Analysis Center/California Institute of Technology, funded by the National Aeronautics and Space Administration and the National Science Foundation. 

This publication makes use of data products from the Wide-field Infrared Survey Explorer, which is a joint project of the University of California, Los Angeles, and the Jet Propulsion Laboratory/California Institute of Technology, funded by the National Aeronautics and Space Administration. 

This paper includes data gathered with the 6.5 meter Magellan Telescopes located at Las Campanas Observatory, Chile.

This work made use of Astrocook \citep[][]{astrocook}, \texttt{Astropy} \citep[][]{astropy1, astropy2, astropy3}, \texttt{IPython} \citep[][]{ipython}, the Julia programming language \citep[][]{Julia}, \texttt{Jupyter Notebooks} \citep[][]{Jupyter}, \texttt{matplotlib} \citep[][]{matplotlib}, \texttt{Numpy} \citep[][]{numpy}, \texttt{Sklearn} \citep[][]{sklearn}, and \texttt{tqdm} \citep[][]{tqdm}.

\section*{Data Availability}
The data underlying this article will be shared on reasonable request to the corresponding author.



\bibliographystyle{mnras}
\bibliography{bib} 

\begin{thebibliography}{}
\makeatletter
\relax
\def\mn@urlcharsother{\let\do\@makeother \do\$\do\&\do\#\do\^\do\_\do\%\do\~}
\def\mn@doi{\begingroup\mn@urlcharsother \@ifnextchar [ {\mn@doi@}
  {\mn@doi@[]}}
\def\mn@doi@[#1]#2{\def\@tempa{#1}\ifx\@tempa\@empty \href
  {http://dx.doi.org/#2} {doi:#2}\else \href {http://dx.doi.org/#2} {#1}\fi
  \endgroup}
\def\mn@eprint#1#2{\mn@eprint@#1:#2::\@nil}
\def\mn@eprint@arXiv#1{\href {http://arxiv.org/abs/#1} {{\tt arXiv:#1}}}
\def\mn@eprint@dblp#1{\href {http://dblp.uni-trier.de/rec/bibtex/#1.xml}
  {dblp:#1}}
\def\mn@eprint@#1:#2:#3:#4\@nil{\def\@tempa {#1}\def\@tempb {#2}\def\@tempc
  {#3}\ifx \@tempc \@empty \let \@tempc \@tempb \let \@tempb \@tempa \fi \ifx
  \@tempb \@empty \def\@tempb {arXiv}\fi \@ifundefined
  {mn@eprint@\@tempb}{\@tempb:\@tempc}{\expandafter \expandafter \csname
  mn@eprint@\@tempb\endcsname \expandafter{\@tempc}}}

\bibitem[\protect\citeauthoryear{Anderson}{Anderson}{2003}]{ref:CCA}
Anderson T.~W.,  2003, An Introduction to Multivariate Statistical Analysis, 3
  edn.
Wiley series in probability and mathematical statistics, Wiley, \url
  {https://www.wiley.com/en-us/An+Introduction+to+Multivariate+Statistical+Analysis%2C+3rd+Edition-p-9780471360919}

\bibitem[\protect\citeauthoryear{{Astropy Collaboration} et~al.,}{{Astropy
  Collaboration} et~al.}{2013}]{astropy1}
{Astropy Collaboration} et~al., 2013, \mn@doi [\aap]
  {10.1051/0004-6361/201322068}, \href
  {https://ui.adsabs.harvard.edu/abs/2013A&A...558A..33A} {558, A33}

\bibitem[\protect\citeauthoryear{{Astropy Collaboration} et~al.,}{{Astropy
  Collaboration} et~al.}{2018}]{astropy2}
{Astropy Collaboration} et~al., 2018, \mn@doi [\aj] {10.3847/1538-3881/aabc4f},
  \href {https://ui.adsabs.harvard.edu/abs/2018AJ....156..123A} {156, 123}

\bibitem[\protect\citeauthoryear{{Astropy Collaboration} et~al.,}{{Astropy
  Collaboration} et~al.}{2022}]{astropy3}
{Astropy Collaboration} et~al., 2022, \mn@doi [\apj]
  {10.3847/1538-4357/ac7c74}, \href
  {https://ui.adsabs.harvard.edu/abs/2022ApJ...935..167A} {935, 167}

\bibitem[\protect\citeauthoryear{{Baron}}{{Baron}}{2019}]{Baron19}
{Baron} D.,  2019, arXiv e-prints, \href
  {https://ui.adsabs.harvard.edu/abs/2019arXiv190407248B} {p. arXiv:1904.07248}

\bibitem[\protect\citeauthoryear{Bezanson, Edelman, Karpinski  \&
  Shah}{Bezanson et~al.}{2017}]{Julia}
Bezanson J.,  Edelman A.,  Karpinski S.,   Shah V.~B.,  2017, \mn@doi [SIAM
  {R}eview] {10.1137/141000671}, 59, 65

\bibitem[\protect\citeauthoryear{{Bianchi} \& {Shiao}}{{Bianchi} \&
  {Shiao}}{2020}]{BianchiGALEX:2020ApJS..250...36B}
{Bianchi} L.,  {Shiao} B.,  2020, \mn@doi [\apjs] {10.3847/1538-4365/aba2d7},
  \href {https://ui.adsabs.harvard.edu/abs/2020ApJS..250...36B} {250, 36}

\bibitem[\protect\citeauthoryear{{Bianchini}, {Fabbian}, {Lapi},
  {Gonzalez-Nuevo}, {Gilli}  \& {Baccigalupi}}{{Bianchini}
  et~al.}{2019}]{Bianchini2019}
{Bianchini} F.,  {Fabbian} G.,  {Lapi} A.,  {Gonzalez-Nuevo} J.,  {Gilli} R.,
  {Baccigalupi} C.,  2019, \mn@doi [\apj] {10.3847/1538-4357/aaf86b}, \href
  {https://ui.adsabs.harvard.edu/abs/2019ApJ...871..136B} {871, 136}

\bibitem[\protect\citeauthoryear{{Boutsia} et~al.,}{{Boutsia}
  et~al.}{2020}]{Boutsia20:2020ApJS..250...26B}
{Boutsia} K.,  et~al., 2020, \mn@doi [\apjs] {10.3847/1538-4365/abafc1}, \href
  {https://ui.adsabs.harvard.edu/abs/2020ApJS..250...26B} {250, 26}

\bibitem[\protect\citeauthoryear{{Boutsia} et~al.,}{{Boutsia}
  et~al.}{2021}]{LF_Boutsia:2021ApJ...912..111B}
{Boutsia} K.,  et~al., 2021, \mn@doi [\apj] {10.3847/1538-4357/abedb5}, \href
  {https://ui.adsabs.harvard.edu/abs/2021ApJ...912..111B} {912, 111}

\bibitem[\protect\citeauthoryear{{Breiman}}{{Breiman}}{2001}]{ref:BreimanRF}
{Breiman} L.,  2001, \mn@doi [Machine Learning] {10.1023/A:1010933404324},
  \href {https://ui.adsabs.harvard.edu/abs/2001MachL..45....5B} {45, 5}

\bibitem[\protect\citeauthoryear{{Calderone} et~al.,}{{Calderone}
  et~al.}{2019}]{Calderone19:2019ApJ...887..268C}
{Calderone} G.,  et~al., 2019, \mn@doi [\apj] {10.3847/1538-4357/ab510a}, \href
  {https://ui.adsabs.harvard.edu/abs/2019ApJ...887..268C} {887, 268}

\bibitem[\protect\citeauthoryear{{Chambers} et~al.,}{{Chambers}
  et~al.}{2016}]{Panstarrs}
{Chambers} K.~C.,  et~al., 2016, arXiv e-prints, \href
  {https://ui.adsabs.harvard.edu/abs/2016arXiv161205560C} {p. arXiv:1612.05560}

\bibitem[\protect\citeauthoryear{{Colless} et~al.,}{{Colless}
  et~al.}{2001}]{2df:2001MNRAS.328.1039C}
{Colless} M.,  et~al., 2001, \mn@doi [\mnras]
  {10.1046/j.1365-8711.2001.04902.x}, \href
  {https://ui.adsabs.harvard.edu/abs/2001MNRAS.328.1039C} {328, 1039}

\bibitem[\protect\citeauthoryear{{Cooke}, {Pettini}, {Jorgenson}, {Murphy}  \&
  {Steidel}}{{Cooke} et~al.}{2014}]{CookePettini14Deuterium}
{Cooke} R.~J.,  {Pettini} M.,  {Jorgenson} R.~A.,  {Murphy} M.~T.,   {Steidel}
  C.~C.,  2014, \mn@doi [\apj] {10.1088/0004-637X/781/1/31}, \href
  {https://ui.adsabs.harvard.edu/abs/2014ApJ...781...31C} {781, 31}

\bibitem[\protect\citeauthoryear{{Cooke}, {Pettini}  \& {Steidel}}{{Cooke}
  et~al.}{2017}]{CookePettini17Deuterium}
{Cooke} R.~J.,  {Pettini} M.,   {Steidel} C.~C.,  2017, \mn@doi [\mnras]
  {10.1093/mnras/stx037}, \href
  {https://ui.adsabs.harvard.edu/abs/2017MNRAS.467..802C} {467, 802}

\bibitem[\protect\citeauthoryear{Cupani, D{\textquotesingle}Odorico, Cristiani,
  Russo, Calderone  \& Taffoni}{Cupani et~al.}{2020}]{astrocook}
Cupani G.,  D{\textquotesingle}Odorico V.,  Cristiani S.,  Russo S.~A.,
  Calderone G.,   Taffoni G.,  2020, in Guzman J.~C.,  Ibsen J.,  eds, Software
  and Cyberinfrastructure for Astronomy {VI}. {SPIE},
  \mn@doi{10.1117/12.2561343}, \url {https://doi.org/10.1117/12.2561343}

\bibitem[\protect\citeauthoryear{{Cupani} et~al.,}{{Cupani}
  et~al.}{2022}]{ref:cupaniFeLoBALs}
{Cupani} G.,  et~al., 2022, \mn@doi [\mnras] {10.1093/mnras/stab3562}, \href
  {https://ui.adsabs.harvard.edu/abs/2022MNRAS.510.2509C} {510, 2509}

\bibitem[\protect\citeauthoryear{{European Southern Observatory}}{{European
  Southern Observatory}}{2013}]{ref:midas}
{European Southern Observatory} 2013, {ESO-MIDAS: General tools for image
  processing and data reduction}, Astrophysics Source Code Library, record
  ascl:1302.017 (\mn@eprint {ascl} {1302.017})

\bibitem[\protect\citeauthoryear{{Fontanot}, {Cristiani}  \&
  {Vanzella}}{{Fontanot} et~al.}{2012}]{FontanotReionization12}
{Fontanot} F.,  {Cristiani} S.,   {Vanzella} E.,  2012, \mn@doi [\mnras]
  {10.1111/j.1365-2966.2012.21594.x}, \href
  {https://ui.adsabs.harvard.edu/abs/2012MNRAS.425.1413F} {425, 1413}

\bibitem[\protect\citeauthoryear{{Fontanot} et~al.,}{{Fontanot}
  et~al.}{2020}]{FontanotSMBHEvo:2020MNRAS.496.3943F}
{Fontanot} F.,  et~al., 2020, \mn@doi [\mnras] {10.1093/mnras/staa1716}, \href
  {https://ui.adsabs.harvard.edu/abs/2020MNRAS.496.3943F} {496, 3943}

\bibitem[\protect\citeauthoryear{{Gaia Collaboration} et~al.,}{{Gaia
  Collaboration} et~al.}{2021}]{GaiaEDr3:2021A&A...649A...1G}
{Gaia Collaboration} et~al., 2021, \mn@doi [\aap]
  {10.1051/0004-6361/202039657}, \href
  {https://ui.adsabs.harvard.edu/abs/2021A&A...649A...1G} {649, A1}

\bibitem[\protect\citeauthoryear{{Grazian} et~al.,}{{Grazian}
  et~al.}{2022}]{LF_Hz_Grazian:2021arXiv211013736G}
{Grazian} A.,  et~al., 2022, \mn@doi [\apj] {10.3847/1538-4357/ac33a4}, \href
  {https://ui.adsabs.harvard.edu/abs/2022ApJ...924...62G} {924, 62}

\bibitem[\protect\citeauthoryear{{Guarneri}, {Calderone}, {Cristiani},
  {Fontanot}, {Boutsia}, {Cupani}, {Grazian}  \& {D'Odorico}}{{Guarneri}
  et~al.}{2021}]{Guarneri:2021MNRAS.506.2471G}
{Guarneri} F.,  {Calderone} G.,  {Cristiani} S.,  {Fontanot} F.,  {Boutsia} K.,
   {Cupani} G.,  {Grazian} A.,   {D'Odorico} V.,  2021, \mn@doi [\mnras]
  {10.1093/mnras/stab1867}, \href
  {https://ui.adsabs.harvard.edu/abs/2021MNRAS.506.2471G} {506, 2471}

\bibitem[\protect\citeauthoryear{Harris et~al.,}{Harris et~al.}{2020}]{numpy}
Harris C.~R.,  et~al., 2020, \mn@doi [Nature] {10.1038/s41586-020-2649-2}, 585,
  357

\bibitem[\protect\citeauthoryear{Hunter}{Hunter}{2007}]{matplotlib}
Hunter J.~D.,  2007, \mn@doi [Computing in Science \& Engineering]
  {10.1109/MCSE.2007.55}, 9, 90

\bibitem[\protect\citeauthoryear{{Inoue}, {Shimizu}, {Iwata}  \&
  {Tanaka}}{{Inoue} et~al.}{2014}]{Inoue2014}
{Inoue} A.~K.,  {Shimizu} I.,  {Iwata} I.,   {Tanaka} M.,  2014, \mn@doi
  [\mnras] {10.1093/mnras/stu936}, \href
  {https://ui.adsabs.harvard.edu/abs/2014MNRAS.442.1805I} {442, 1805}

\bibitem[\protect\citeauthoryear{Japkowicz \& Stephen}{Japkowicz \&
  Stephen}{2002}]{Paper:UnbalancedJustification:journals/ida/JapkowiczS02}
Japkowicz N.,  Stephen S.,  2002, Intelligent Data Analysis, 6, 429

\bibitem[\protect\citeauthoryear{{Jones} et~al.,}{{Jones}
  et~al.}{2009}]{6df:2009MNRAS.399..683J}
{Jones} D.~H.,  et~al., 2009, \mn@doi [\mnras]
  {10.1111/j.1365-2966.2009.15338.x}, \href
  {https://ui.adsabs.harvard.edu/abs/2009MNRAS.399..683J} {399, 683}

\bibitem[\protect\citeauthoryear{Kluyver et~al.,}{Kluyver
  et~al.}{2016}]{Jupyter}
Kluyver T.,  et~al., 2016, in Loizides F.,  Schmidt B.,  eds, Positioning and
  Power in Academic Publishing: Players, Agents and Agendas. pp 87 -- 90

\bibitem[\protect\citeauthoryear{{Krawczyk}, {Richards}, {Mehta}, {Vogeley},
  {Gallagher}, {Leighly}, {Ross}  \& {Schneider}}{{Krawczyk}
  et~al.}{2013}]{Krawczyk2013}
{Krawczyk} C.~M.,  {Richards} G.~T.,  {Mehta} S.~S.,  {Vogeley} M.~S.,
  {Gallagher} S.~C.,  {Leighly} K.~M.,  {Ross} N.~P.,   {Schneider} D.~P.,
  2013, \mn@doi [\apjs] {10.1088/0067-0049/206/1/4}, \href
  {https://ui.adsabs.harvard.edu/abs/2013ApJS..206....4K} {206, 4}

\bibitem[\protect\citeauthoryear{{Kulkarni}, {Worseck}  \&
  {Hennawi}}{{Kulkarni} et~al.}{2019}]{Kulkarni19}
{Kulkarni} G.,  {Worseck} G.,   {Hennawi} J.~F.,  2019, \mn@doi [\mnras]
  {10.1093/mnras/stz1493}, \href
  {https://ui.adsabs.harvard.edu/abs/2019MNRAS.488.1035K} {488, 1035}

\bibitem[\protect\citeauthoryear{{Laor} \& {Netzer}}{{Laor} \&
  {Netzer}}{1989}]{Laor1989}
{Laor} A.,  {Netzer} H.,  1989, \mn@doi [\mnras] {10.1093/mnras/238.3.897},
  \href {https://ui.adsabs.harvard.edu/abs/1989MNRAS.238..897L} {238, 897}

\bibitem[\protect\citeauthoryear{{Liske} et~al.,}{{Liske}
  et~al.}{2008}]{Liske+08:2008MNRAS.386.1192L}
{Liske} J.,  et~al., 2008, \mn@doi [\mnras] {10.1111/j.1365-2966.2008.13090.x},
  \href {https://ui.adsabs.harvard.edu/abs/2008MNRAS.386.1192L} {386, 1192}

\bibitem[\protect\citeauthoryear{{Lyke} et~al.,}{{Lyke}
  et~al.}{2020}]{LykeSDSS16q:2020ApJS..250....8L}
{Lyke} B.~W.,  et~al., 2020, \mn@doi [\apjs] {10.3847/1538-4365/aba623}, \href
  {https://ui.adsabs.harvard.edu/abs/2020ApJS..250....8L} {250, 8}

\bibitem[\protect\citeauthoryear{{Milakovi{\'c}}, {Lee}, {Carswell}, {Webb},
  {Molaro}  \& {Pasquini}}{{Milakovi{\'c}}
  et~al.}{2021}]{FundConstMilakovic:2021MNRAS.500....1M}
{Milakovi{\'c}} D.,  {Lee} C.-C.,  {Carswell} R.~F.,  {Webb} J.~K.,  {Molaro}
  P.,   {Pasquini} L.,  2021, \mn@doi [\mnras] {10.1093/mnras/staa3217}, \href
  {https://ui.adsabs.harvard.edu/abs/2021MNRAS.500....1M} {500, 1}

\bibitem[\protect\citeauthoryear{{Mor}, {Netzer}  \& {Elitzur}}{{Mor}
  et~al.}{2009}]{Mor2009}
{Mor} R.,  {Netzer} H.,   {Elitzur} M.,  2009, \mn@doi [\apj]
  {10.1088/0004-637X/705/1/298}, \href
  {https://ui.adsabs.harvard.edu/abs/2009ApJ...705..298M} {705, 298}

\bibitem[\protect\citeauthoryear{{Murphy} et~al.,}{{Murphy}
  et~al.}{2022}]{FundConstMurphy:ApJ}
{Murphy} M.~T.,  et~al., 2022, \mn@doi [\aap] {10.1051/0004-6361/202142257},
  \href {https://ui.adsabs.harvard.edu/abs/2022A&A...658A.123M} {658, A123}

\bibitem[\protect\citeauthoryear{{Nakoneczny} et~al.,}{{Nakoneczny}
  et~al.}{2021}]{Nakoneczny21:2021A&A...649A..81N}
{Nakoneczny} S.~J.,  et~al., 2021, \mn@doi [\aap]
  {10.1051/0004-6361/202039684}, \href
  {https://ui.adsabs.harvard.edu/abs/2021A&A...649A..81N} {649, A81}

\bibitem[\protect\citeauthoryear{{Nikolenko}}{{Nikolenko}}{2021}]{Nikolenko2021}
{Nikolenko} S.,  2021, \mn@doi [Synthetic Data for Deep Learning, Springer]
  {https://doi.org/10.1007/978-3-030-75178-4}

\bibitem[\protect\citeauthoryear{{Onken} et~al.,}{{Onken}
  et~al.}{2019}]{SkyMapper3:2019PASA...36...33O}
{Onken} C.~A.,  et~al., 2019, \mn@doi [\pasa] {10.25914/5f14eded2d116}, \href
  {https://ui.adsabs.harvard.edu/abs/2019PASA...36...33O} {36, e033}

\bibitem[\protect\citeauthoryear{{Onken}, {Wolf}, {Bian}, {Fan}, {Jeat Hon},
  {Raithel}, {Tisserand}  \& {Lai}}{{Onken}
  et~al.}{2021}]{Onken21:2021arXiv210512215O}
{Onken} C.~A.,  {Wolf} C.,  {Bian} F.,  {Fan} X.,  {Jeat Hon} W.,  {Raithel}
  D.,  {Tisserand} P.,   {Lai} S.,  2021, arXiv e-prints, \href
  {https://ui.adsabs.harvard.edu/abs/2021arXiv210512215O} {p. arXiv:2105.12215}

\bibitem[\protect\citeauthoryear{Pedregosa et~al.,}{Pedregosa
  et~al.}{2011}]{sklearn}
Pedregosa F.,  et~al., 2011, Journal of Machine Learning Research, 12, 2825

\bibitem[\protect\citeauthoryear{P\'erez \& Granger}{P\'erez \&
  Granger}{2007}]{ipython}
P\'erez F.,  Granger B.~E.,  2007, \mn@doi [Computing in Science and
  Engineering] {10.1109/MCSE.2007.53}, 9, 21

\bibitem[\protect\citeauthoryear{{Pier} \& {Krolik}}{{Pier} \&
  {Krolik}}{1993}]{Pier1993}
{Pier} E.~A.,  {Krolik} J.~H.,  1993, \mn@doi [\apj] {10.1086/173427}, \href
  {https://ui.adsabs.harvard.edu/abs/1993ApJ...418..673P} {418, 673}

\bibitem[\protect\citeauthoryear{{Prochaska}}{{Prochaska}}{2017}]{specDB}
{Prochaska} J.~X.,  2017, \mn@doi [Astronomy and Computing]
  {10.1016/j.ascom.2017.03.003}, \href
  {https://ui.adsabs.harvard.edu/abs/2017A&C....19...27P} {19, 27}

\bibitem[\protect\citeauthoryear{{Reis}, {Baron}  \& {Shahaf}}{{Reis}
  et~al.}{2019}]{ReisPRF:2019AJ....157...16R}
{Reis} I.,  {Baron} D.,   {Shahaf} S.,  2019, \mn@doi [\aj]
  {10.3847/1538-3881/aaf101}, \href
  {https://ui.adsabs.harvard.edu/abs/2019AJ....157...16R} {157, 16}

\bibitem[\protect\citeauthoryear{{Richards} et~al.,}{{Richards}
  et~al.}{2002}]{SDSSQSOvsStellarColors}
{Richards} G.~T.,  et~al., 2002, \mn@doi [\aj] {10.1086/340187}, \href
  {https://ui.adsabs.harvard.edu/abs/2002AJ....123.2945R} {123, 2945}

\bibitem[\protect\citeauthoryear{{Richards} et~al.,}{{Richards}
  et~al.}{2006}]{Richards2006}
{Richards} G.~T.,  et~al., 2006, \mn@doi [\apjs] {10.1086/506525}, \href
  {https://ui.adsabs.harvard.edu/abs/2006ApJS..166..470R} {166, 470}

\bibitem[\protect\citeauthoryear{{Schindler} et~al.,}{{Schindler}
  et~al.}{2019a}]{SDSSIncomplete_Schindler_PS:2019ApJ...871..258S}
{Schindler} J.-T.,  et~al., 2019a, \mn@doi [\apjs] {10.3847/1538-4365/ab20d0},
  \href {https://ui.adsabs.harvard.edu/abs/2019ApJS..243....5S} {243, 5}

\bibitem[\protect\citeauthoryear{{Schindler} et~al.,}{{Schindler}
  et~al.}{2019b}]{SDSSIncomplete_Schindler:2019ApJ...871..258S}
{Schindler} J.-T.,  et~al., 2019b, \mn@doi [\apj] {10.3847/1538-4357/aaf86c},
  \href {https://ui.adsabs.harvard.edu/abs/2019ApJ...871..258S} {871, 258}

\bibitem[\protect\citeauthoryear{{Sevilla-Noarbe} et~al.,}{{Sevilla-Noarbe}
  et~al.}{2021}]{DESY3Gold}
{Sevilla-Noarbe} I.,  et~al., 2021, \mn@doi [\apjs] {10.3847/1538-4365/abeb66},
  \href {https://ui.adsabs.harvard.edu/abs/2021ApJS..254...24S} {254, 24}

\bibitem[\protect\citeauthoryear{{Skrutskie} et~al.,}{{Skrutskie}
  et~al.}{2006}]{2MASS:2006AJ....131.1163S}
{Skrutskie} M.~F.,  et~al., 2006, \mn@doi [\aj] {10.1086/498708}, \href
  {https://ui.adsabs.harvard.edu/abs/2006AJ....131.1163S} {131, 1163}

\bibitem[\protect\citeauthoryear{{Sun} \& {Malkan}}{{Sun} \&
  {Malkan}}{1989}]{Sun1989}
{Sun} W.-H.,  {Malkan} M.~A.,  1989, \mn@doi [\apj] {10.1086/167986}, \href
  {https://ui.adsabs.harvard.edu/abs/1989ApJ...346...68S} {346, 68}

\bibitem[\protect\citeauthoryear{{Vanden Berk} et~al.,}{{Vanden Berk}
  et~al.}{2001}]{VandenBerk2001}
{Vanden Berk} D.~E.,  et~al., 2001, \mn@doi [\aj] {10.1086/321167}, \href
  {https://ui.adsabs.harvard.edu/abs/2001AJ....122..549V} {122, 549}

\bibitem[\protect\citeauthoryear{{V{\'e}ron-Cetty} \&
  {V{\'e}ron}}{{V{\'e}ron-Cetty} \&
  {V{\'e}ron}}{2010}]{Veron10:2010A&A...518A..10V}
{V{\'e}ron-Cetty} M.~P.,  {V{\'e}ron} P.,  2010, \mn@doi [\aap]
  {10.1051/0004-6361/201014188}, \href
  {https://ui.adsabs.harvard.edu/abs/2010A&A...518A..10V} {518, A10}

\bibitem[\protect\citeauthoryear{{Wenzl} et~al.,}{{Wenzl}
  et~al.}{2021}]{QuasarSelection1:2021AJ....162...72W}
{Wenzl} L.,  et~al., 2021, \mn@doi [\aj] {10.3847/1538-3881/ac0254}, \href
  {https://ui.adsabs.harvard.edu/abs/2021AJ....162...72W} {162, 72}

\bibitem[\protect\citeauthoryear{{Wolf} et~al.,}{{Wolf}
  et~al.}{2020}]{Wolf20QSO:2020MNRAS.491.1970W}
{Wolf} C.,  et~al., 2020, \mn@doi [\mnras] {10.1093/mnras/stz2955}, \href
  {https://ui.adsabs.harvard.edu/abs/2020MNRAS.491.1970W} {491, 1970}

\bibitem[\protect\citeauthoryear{{Wright} et~al.,}{{Wright}
  et~al.}{2010}]{WISE:2010AJ....140.1868W}
{Wright} E.~L.,  et~al., 2010, \mn@doi [\aj] {10.1088/0004-6256/140/6/1868},
  \href {https://ui.adsabs.harvard.edu/abs/2010AJ....140.1868W} {140, 1868}

\bibitem[\protect\citeauthoryear{{Wu} et~al.,}{{Wu}
  et~al.}{2015}]{2015Natur.518..512W}
{Wu} X.-B.,  et~al., 2015, \mn@doi [\nat] {10.1038/nature14241}, \href
  {https://ui.adsabs.harvard.edu/abs/2015Natur.518..512W} {518, 512}

\bibitem[\protect\citeauthoryear{da Costa-Luis et~al.,}{da~Costa-Luis
  et~al.}{2022}]{tqdm}
da Costa-Luis C.,  et~al., 2022, {tqdm: A fast, Extensible Progress Bar for
  Python and CLI}, \mn@doi{10.5281/zenodo.7046742}, \url
  {https://doi.org/10.5281/zenodo.7046742}

\makeatother
\end{thebibliography}

\appendix

\onecolumn
\section{Spectroscopic catalogue}

\begin{longtable}{ccccccccc}
\caption{Observed sources with a reliable spectroscopic identification. The quoted $i_{\rm psf}$ magnitude is in the AB photometric system, and targets identified with * show broad absorption lines (BAL QSOs).}
\label{tab:NewSpec} \\
\toprule
     & QID       & RA            & DEC     & $i_{\rm psf}$ & z$_{\rm spec}$ & Class & Obs. date & Instrument \\
     &           & J2000         & J2000         &           &            &       &           &  \\
\midrule
1	& 811621	& 11:22:35.81	& --07:33:38.68	& 17.822	& 2.453	& QSO	    & 2021--04--16    & DOLORES \\
2	& 812649*	& 12:20:08.04	& --03:49:43.99	& 16.759	& 3.117	& QSO	    & 2021--04--20    & LDSS3 \\
3	& 813281	& 23:34:45.77	& --12:20:20.34	& 17.801	& 2.896	& QSO	    & 2021--10--26    & DOLORES \\
4	& 819732	& 00:24:23.76	& --14:25:02.75	& 17.953	& 2.328	& QSO	    & 2021--10--11    & DOLORES \\
5	& 822502	& 11:34:51.50	& +02:02:08.57	& 17.724	& 2.621	& QSO	    & 2021--04--08    & DOLORES \\
6	& 823202	& 13:18:33.31	& --02:45:36.22	& 17.888	& 1.404	& QSO	    & 2021--06--25    & DOLORES \\
7	& 824842	& 00:00:47.92	& --17:52:39.51	& 17.940    & 1.784	& QSO	    & 2021--10--11    & DOLORES \\
8	& 827601	& 23:02:11.26	& --01:34:39.41	& 17.835	& 2.571	& QSO	    & 2021--08--13    & DOLORES \\
9	& 827911	& 00:35:37.25	& --05:46:55.64	& 17.945	& 2.908	& QSO	    & 2021--08--13    & DOLORES \\
10	& 828680	& 13:52:37.23	& --10:12:03.90	& 17.454	& 3.275	& QSO	    & 2021--04--17    & LDSS3 \\
11	& 829059	& 14:07:01.79	& --24:15:02.33	& 17.713	& 3.057	& QSO	    & 2021--04--17    & LDSS3 \\
12	& 830473	& 13:47:41.79	& --08:38:43.28	& 17.978	& 3.205	& QSO	    & 2021--04--17    & LDSS3 \\
13	& 830592*	& 14:05:29.04	& --14:05:13.92	& 17.650    & 3.094	& QSO	    & 2021--04--08    & DOLORES \\
14	& 830694	& 13:58:35.67	& --20:22:14.11	& 17.809	& 2.873	& QSO	    & 2021--06--01    & LDSS3 \\
15	& 832405	& 13:37:51.03	& --07:42:33.39	& 17.654	& 3.07	& QSO	    & 2021--04--08    & DOLORES \\
16	& 839684	& 15:28:45.83	& --22:57:18.32	& 17.728	& 2.82	& QSO	    & 2021--04--17    & LDSS3 \\
17	& 841855	& 19:07:51.48	& --66:06:18.04	& 17.672	& 3.136	& QSO	    & 2021--05--31    & LDSS3 \\
18	& 846269	& 23:05:17.49	& --73:05:29.09	& 17.997	& 2.704	& QSO	    & 2021--10--24    & LDSS3 \\
19	& 847891	& 04:51:46.91	& --30:44:09.03	& 17.956	& 3.354	& QSO	    & 2021--10--24    & LDSS3 \\
20	& 848374	& 03:27:03.00	& --20:10:36.99	& 17.903	& 3.065	& QSO	    & 2021--10--24    & LDSS3 \\
21	& 852752	& 05:29:07.26	& --39:58:44.88	& 17.745	& 3.058	& QSO	    & 2021--10--23    & LDSS3 \\
22	& 853838	& 03:32:15.75	& --16:59:04.10	& 17.885	& 2.44	& QSO	    & 2021--10--28    & DOLORES \\
23	& 853947	& 03:36:41.63	& --15:27:07.90	& 17.939	& 2.673	& QSO	    & 2021--10--28    & DOLORES \\
24	& 856212*	& 21:52:31.21	& --23:11:15.66	& 17.623	& 2.24	& QSO	    & 2021--07--27    & LDSS3 \\
25	& 856826*	& 22:06:32.08	& --23:45:44.56	& 17.646	& 2.6	& QSO	    & 2021--07--24    & LDSS3 \\
26	& 858022	& 21:02:10.69	& --14:31:01.87	& 17.874	& 2.78	& QSO	    & 2021--08--12    & DOLORES \\
27	& 858180	& 21:57:25.92	& --10:37:22.85	& 17.917	& 3.117	& QSO	    & 2021--08--13    & DOLORES \\
28	& 859971	& 20:55:50.60	& +01:06:15.46	& 17.591	& 0.0	& Star	    & 2021--08--12    & DOLORES \\
29	& 864194	& 23:18:06.56	& --70:28:00.80	& 17.535	& 2.006	& QSO	    & 2021--10--23    & LDSS3 \\
30	& 864319	& 23:04:13.58	& --57:16:12.66	& 17.778	& 3.221	& QSO	    & 2021--11--14    & EFOSC \\
31	& 869710	& 21:06:31.89	& --18:45:32.46	& 17.862	& 2.157	& QSO	    & 2021--08--12    & DOLORES \\
32	& 872817	& 22:05:46.02	& --51:58:19.33	& 17.289	& 2.847	& QSO	    & 2021--07--25    & LDSS3 \\
33	& 873739	& 21:46:15.35	& --39:38:34.73	& 17.688	& 2.773	& QSO	    & 2021--07--25    & LDSS3 \\
34	& 875386	& 21:35:11.99	& --27:50:19.80	& 17.928	& 2.372	& QSO	    & 2021--07--27    & LDSS3 \\
35	& 877355	& 15:05:59.48	& --08:34:39.71	& 17.935	& 3.012	& QSO	    & 2021--08--12    & DOLORES \\
36	& 877393	& 15:11:32.87	& --07:20:56.31	& 17.966	& 2.631	& QSO	    & 2021--08--12    & DOLORES \\
37	& 879338	& 15:32:08.06	& --06:13:59.60	& 17.468	& 3.546	& QSO	    & 2021--04--16    & LDSS3 \\
38	& 889097*	& 00:10:11.97	& --66:26:55.08	& 17.961	& 2.347	& QSO	    & 2021--11--14    & EFOSC \\
39	& 891070	& 22:16:01.07	& --31:24:53.43	& 17.938	& 2.442	& QSO	    & 2021--11--16    & EFOSC \\
40	& 891397	& 23:00:54.63	& --54:16:17.68	& 17.887	& 3.966	& QSO	    & 2021--11--15    & EFOSC \\
41	& 901267	& 11:11:50.60	& --12:15:54.00	& 17.792	& 2.531	& QSO	    & 2021--04--16    & DOLORES \\
42	& 903300	& 10:46:50.77	& --19:13:36.82	& 17.767	& 2.055	& QSO	    & 2021--04--16    & DOLORES \\
43	& 904188	& 11:03:33.29	& --06:49:52.05	& 17.705	& 2.916	& QSO	    & 2021--04--08    & DOLORES \\
44	& 905520	& 01:29:35.26	& --44:42:22.66	& 17.973	& 2.327	& QSO	    & 2021--10--23    & LDSS3 \\
45	& 913533	& 10:19:20.03	& --19:20:40.17	& 17.916	& 2.603	& QSO	    & 2021--04--07    & DOLORES \\
46	& 915280	& 10:30:57.17	& --10:11:02.80	& 17.595	& 2.421	& QSO	    & 2021--04--16    & DOLORES \\
47	& 917210*	& 22:29:38.43	& --55:20:50.07	& 17.758	& 2.749	& QSO	    & 2021--09--23    & LDSS3 \\
48	& 917871	& 21:05:57.21	& --52:59:40.17	& 17.691	& 2.424	& QSO	    & 2021--07--27    & LDSS3 \\
49	& 919191	& 23:19:56.33	& --48:21:48.70	& 17.665	& 2.695	& QSO	    & 2021--10--24    & LDSS3 \\
50	& 938885	& 02:44:32.26	& --57:33:34.85	& 17.884	& 3.238	& QSO	    & 2021--10--24    & LDSS3 \\
51	& 951660	& 01:19:31.63	& --51:11:45.41	& 17.842	& 3.586	& QSO	    & 2021--10--23    & LDSS3 \\
52	& 954876	& 01:49:24.76	& --72:04:32.34	& 17.828	& 1.899	& QSO	    & 2021--10--24    & LDSS3 \\
53	& 956571	& 00:36:48.87	& --48:59:38.15	& 17.960    & 3.046	& QSO	    & 2021--10--23    & LDSS3 \\
54	& 958147	& 00:56:44.68	& --48:19:03.50	& 17.973	& 3.201	& QSO	    & 2021--10--24    & LDSS3 \\
55	& 959812*	& 04:34:30.93	& --80:37:24.94	& 17.865	& 2.611	& QSO	    & 2021--10--23    & LDSS3 \\
56	& 960673	& 21:16:29.48	& --65:20:26.98	& 17.987	& 3.399	& QSO	    & 2021--11--14    & EFOSC \\
57	& 961314	& 21:03:45.25	& --44:43:34.14	& 17.957	& 2.651	& QSO	    & 2021--10--24    & LDSS3 \\
58	& 963893	& 21:26:52.27	& --51:00:42.24	& 17.852	& 3.512	& QSO	    & 2021--09--23    & LDSS3 \\
59	& 968067	& 20:49:21.09	& --16:43:38.43	& 17.986	& 2.096	& QSO	    & 2021--10--10    & DOLORES \\
60	& 968243	& 21:32:06.44	& --15:46:01.93	& 17.440    & 1.817	& QSO	    & 2021--08--13    & DOLORES \\
61	& 971720	& 13:23:51.36	& --37:11:28.04	& 17.158	& 0.147	& Galaxy	& 2021--04--16    & LDSS3 \\
62	& 977267	& 22:54:48.84	& --64:30:11.41	& 17.922	& 0.048	& Galaxy	& 2021--10--23    & LDSS3 \\
63	& 979960	& 22:18:51.08	& --61:50:43.63	& 17.760    & 3.315	& QSO	    & 2021--07--27    & LDSS3 \\
64	& 982938	& 22:25:09.49	& --47:10:06.50	& 17.851	& 2.706	& QSO	    & 2021--07--27    & LDSS3 \\
65	& 984341	& 14:24:10.85	& --30:07:10.37	& 17.829	& 2.044	& QSO	    & 2021--06--01    & LDSS3 \\
66	& 995617	& 23:07:37.73	& --75:36:23.48	& 17.848	& 2.059	& QSO	    & 2021--10--23    & LDSS3 \\
67	& 996288*	& 18:51:16.68	& --76:26:28.59	& 17.934	& 2.466	& QSO	    & 2021--06--01    & LDSS3 \\
68	& 997809	& 15:44:25.71	& --13:03:48.93	& 17.897	& 3.226	& QSO	    & 2021--04--17    & LDSS3 \\
69	& 999385	& 05:16:43.80	& --39:31:21.89	& 17.885	& 2.916	& QSO	    & 2021--10--23    & LDSS3 \\
70	& 1006739*	& 15:08:13.52	& --16:46:34.71	& 17.970    & 2.076	& QSO	    & 2021--04--17    & LDSS3 \\
71	& 1007178	& 14:32:46.53	& --10:59:44.47	& 17.861	& 2.634	& QSO	    & 2021--04--17    & LDSS3 \\
72	& 1007867	& 04:31:07.32	& --09:14:11.19	& 17.599	& 3.698	& QSO	    & 2021--11--17    & EFOSC \\
73	& 1011659	& 05:28:09.48	& --27:21:20.91	& 17.940    & 3.285	& QSO	    & 2021--10--24    & LDSS3 \\
74	& 1011914	& 05:29:52.11	& --24:27:22.27	& 17.426	& 2.229	& QSO	    & 2021--10--23    & LDSS3 \\
75	& 1017251	& 14:15:00.66	& --27:55:49.29	& 17.914	& 2.745	& QSO	    & 2021--06--01    & LDSS3 \\
76	& 1017341*	& 14:29:43.45	& --22:31:08.69	& 17.686	& 2.607	& QSO	    & 2021--06--01    & LDSS3 \\
77	& 1021276*	& 03:04:17.95	& --54:13:47.52	& 17.833	& 2.73	& QSO	    & 2021--09--22    & LDSS3 \\
78. & 1030322	& 23:28:11.75	& --32:31:01.72	& 17.713	& 3.565	& QSO	    & 2021--07--24    & LDSS3 \\
79	& 1030688	& 00:03:26.87	& --32:52:07.54	& 17.850    & 3.808	& QSO	    & 2021--10--23    & LDSS3 \\
80	& 1032484	& 01:01:04.90	& --40:27:52.05	& 17.986	& 2.998	& QSO	    & 2021--10--23    & LDSS3 \\
81	& 1034730	& 00:51:53.31	& --19:08:32.11	& 17.840    & 2.709	& QSO	    & 2021--08--13    & DOLORES \\
82	& 1034896	& 01:26:11.23	& --31:48:41.94	& 17.904	& 4.016	& QSO	    & 2021--10--24    & LDSS3 \\
83	& 1035799	& 00:01:23.40	& --19:35:55.65	& 17.450    & 2.665	& QSO	    & 2021--08--13    & DOLORES \\
84	& 1036444	& 00:34:08.97	& --39:46:23.31	& 17.966	& 2.672	& QSO	    & 2021--10--23    & LDSS3 \\
85	& 1036840	& 23:43:32.11	& --39:23:21.02	& 17.914	& 3.921	& QSO	    & 2021--11--15    & EFOSC \\
86	& 1038071	& 00:07:31.62	& --19:39:00.79	& 17.901	& 2.794	& QSO	    & 2021--10--11    & DOLORES \\
87	& 1040592	& 00:39:26.45	& --13:34:14.68	& 17.963	& 2.453	& QSO	    & 2021--08--13    & DOLORES \\
88	& 1042750	& 23:24:21.54	& --31:34:22.14	& 17.369	& 3.546	& QSO	    & 2021--11--15    & EFOSC \\
89	& 1042794	& 22:51:48.44	& --36:46:21.73	& 17.914	& 3.658	& QSO	    & 2021--10--24    & LDSS3 \\
90	& 1043311	& 22:21:00.86	& +00:22:36.83	& 17.553	& 0.0	& Star	    & 2021--06--18    & DOLORES \\
91	& 1052170	& 19:59:01.20	& --28:26:16.22	& 17.894	& 3.241	& QSO	    & 2021--05--31    & LDSS3 \\
92	& 1052257	& 20:12:14.39	& --22:56:50.50	& 17.948	& 2.7	& QSO	    & 2021--10--23    & LDSS3 \\
93	& 1052311	& 20:16:24.35	& --18:46:25.58	& 17.863	& 3.147	& QSO	    & 2021--08--12    & DOLORES \\
94	& 1059935	& 01:56:08.75	& +02:05:18.75	& 17.764	& 2.917	& QSO	    & 2021--10--27    & DOLORES \\
95	& 1064072	& 02:21:57.12	& --14:14:57.13	& 17.932	& 2.625	& QSO	    & 2021--10--28    & DOLORES \\
96	& 1067635	& 01:41:26.07	& --16:20:22.13	& 17.768	& 2.862	& QSO	    & 2021--08--13    & DOLORES \\
97	& 1068979*	& 02:39:30.71	& --36:29:39.75	& 17.836	& 2.696	& QSO	    & 2021--10--23    & LDSS3 \\
98	& 1069001	& 02:47:10.46	& --22:02:45.68	& 17.814	& 2.798	& QSO	    & 2021--10--24    & LDSS3 \\
99	& 1070075*	& 00:57:22.37	& --38:18:52.10	& 17.774	& 2.389	& QSO	    & 2021--10--23    & LDSS3 \\
100	& 1072238	& 02:52:21.09	& --05:24:25.49	& 17.162	& 3.048	& QSO	    & 2021--10--11    & DOLORES \\
101	& 1076856	& 04:11:31.15	& --34:49:45.45	& 17.808	& 2.792	& QSO	    & 2021--11--14    & EFOSC \\
102	& 1079673	& 03:34:26.48	& --38:20:52.88	& 17.898	& 3.232	& QSO	    & 2021--10--24    & LDSS3 \\
103	& 1086433	& 04:35:13.59	& --44:54:40.89	& 17.845	& 3.217	& QSO	    & 2021--10--23    & LDSS3 \\
104	& 1087276	& 19:54:38.01	& --41:08:31.42	& 17.853	& 2.814	& QSO	    & 2021--06--01    & LDSS3 \\
105	& 1091000	& 03:05:13.35	& --57:49:52.15	& 17.961	& 2.559	& QSO	    & 2021--10--24    & LDSS3 \\
106	& 1106210	& 03:32:18.91	& +01:56:41.00	& 17.920    & 2.163	& QSO	    & 2021--10--11    & DOLORES \\
107	& 1107458	& 05:29:06.69	& --15:46:28.64	& 17.929	& 2.605	& QSO	    & 2021--10--28    & DOLORES \\
108	& 1110306	& 04:42:23.73	& +00:02:17.20	& 17.889	& 2.349	& QSO	    & 2021--10--28    & DOLORES \\
109	& 1111616	& 05:31:27.00	& --49:04:44.00	& 17.890    & 3.207	& QSO	    & 2021--10--23    & LDSS3 \\
110	& 1128997	& 20:17:34.01	& --52:49:46.72	& 17.962	& 3.206	& QSO	    & 2021--10--23    & LDSS3 \\
111	& 1138663	& 20:35:21.75	& --11:45:34.22	& 17.997	& 3.03	& QSO	    & 2021--08--12    & DOLORES \\
112	& 1140469*	& 12:00:09.95	& --29:46:09.76	& 17.951	& 3.068	& QSO	    & 2021--04--16    & LDSS3 \\
113	& 1140534	& 11:37:39.59	& --24:32:53.35	& 17.831	& 3.595	& QSO	    & 2022--02--24    & LDSS3 \\
114	& 1144132	& 05:56:00.65	& --32:43:13.07	& 17.995	& 4.042	& QSO	    & 2021--10--23    & LDSS3 \\
115	& 1150493	& 15:58:04.57	& --04:44:05.87	& 17.951	& 1.52	& QSO	    & 2021--06--18    & DOLORES \\
116	& 1171630	& 09:46:01.67	& --11:05:51.32	& 17.937	& 2.686	& QSO	    & 2021--04--07    & DOLORES \\
117	& 1171778	& 09:48:44.41	& --18:18:32.39	& 17.955	& 3.217	& QSO	    & 2021--04--07    & DOLORES \\
118	& 1179313	& 04:48:09.10	& --22:42:08.97	& 17.934	& 3.421	& QSO	    & 2021--10--24    & LDSS3 \\
119	& 1194166	& 04:53:37.98	& --50:26:20.53	& 17.962	& 3.57	& QSO	    & 2021--09--22    & LDSS3 \\
120	& 1194186	& 04:11:02.50	& --50:13:46.59	& 17.909	& 3.217	& QSO	    & 2021--11--14    & EFOSC \\
121	& 1194925	& 03:06:03.60	& --31:09:44.58	& 17.965	& 2.668	& QSO	    & 2021--11--16    & EFOSC \\
122	& 1202352	& 04:20:00.72	& --52:52:26.75	& 17.998	& 3.695	& QSO	    & 2021--11--14    & EFOSC \\
123	& 1203737*	& 01:03:30.95	& --12:44:20.58	& 17.974	& 3.024	& QSO	    & 2021--10--11    & DOLORES \\
124	& 1203788	& 19:58:03.42	& --30:21:16.17	& 17.846	& 2.741	& QSO	    & 2021--06--01    & LDSS3 \\
125	& 1327620	& 13:37:47.33	& --19:58:11.56	& 16.296	& 0.796	& QSO	    & 2021--04--16    & LDSS3 \\
126	& 1339753	& 17:14:21.39	& +08:58:57.84	& 16.485	& 0.0	& Star	    & 2021--06--18    & DOLORES \\
127	& 1344952	& 13:58:14.92	& --28:06:10.75	& 16.838	& 2.687	& QSO	    & 2021--04--16    & LDSS3 \\
128	& 1351549*	& 21:31:54.88	& +08:07:26.73	& 16.890    & 2.703	& QSO	    & 2021--08--12    & DOLORES \\
129	& 1351687	& 21:43:25.98	& +04:48:34.03	& 16.855	& 0.0	& Star	    & 2021--06--18    & DOLORES \\
130	& 1354768	& 03:54:51.70	& +11:11:14.43	& 16.907	& 0.0	& Star	    & 2021--10--28    & DOLORES \\
131	& 1360069	& 17:04:49.71	& +04:30:16.65	& 16.654	& 0.0	& Star	    & 2021--06--18    & DOLORES \\
132	& 1363142	& 23:04:38.54	& --52:07:09.54	& 17.089	& 2.08	& QSO	    & 2021--10--24    & LDSS3 \\
133	& 1364626	& 02:32:16.87	& +07:09:51.40	& 17.023	& 0.0	& Star	    & 2021--08--13    & DOLORES \\
134	& 1365060	& 10:55:44.02	& +05:16:29.20	& 17.093	& 2.811	& QSO	    & 2021--04--07    & DOLORES \\
135	& 1367993	& 13:52:40.00	& --29:49:57.02	& 17.196	& 2.662	& QSO	    & 2021--04--16    & LDSS3 \\
136	& 1368490	& 15:35:03.14	& +00:45:04.49	& 17.146	& 0.0	& Star	    & 2021--06--18    & DOLORES \\
137	& 1370585	& 02:01:09.17	& +07:20:08.47	& 17.198	& 2.481	& QSO	    & 2021--08--13    & DOLORES \\
138	& 1372270	& 23:38:07.92	& --32:19:05.62	& 17.261	& 2.54	& QSO	    & 2021--09--24    & LDSS3 \\
139	& 1374090	& 11:32:01.79	& --13:23:08.82	& 17.268	& 2.763	& QSO	    & 2021--04--08    & DOLORES \\
140	& 1377402	& 08:53:18.70	& +08:53:24.07	& 17.275	& 2.726	& QSO	    & 2021--11--16    & EFOSC \\
141	& 1380415	& 03:50:00.38	& --18:38:26.54	& 17.322	& 3.107	& QSO	    & 2021--10--28    & DOLORES \\
142	& 1384237	& 04:23:35.74	& +04:11:48.03	& 17.318	& 0.0	& Star	    & 2021--10--28    & DOLORES \\
143	& 1384833	& 13:49:39.42	& +02:00:21.55	& 17.353	& 3.354	& QSO	    & 2021--04--08    & DOLORES \\
144	& 1385296	& 15:20:02.20	& +10:50:55.53	& 17.363	& 2.468	& QSO	    & 2021--04--08    & DOLORES \\
145	& 1386055	& 16:19:05.95	& +06:41:22.68	& 17.407	& 0.0	& Star	    & 2021--08--12    & DOLORES \\
146	& 1386702	& 22:42:49.59	& --05:59:23.31	& 17.415	& 2.666	& QSO	    & 2021--08--13    & DOLORES \\
147	& 1392457	& 12:52:22.75	& --37:48:16.80	& 17.491	& 3.425	& QSO	    & 2021--04--16    & LDSS3 \\
148	& 1393389	& 21:02:17.81	& --17:01:55.08	& 17.509	& 2.837	& QSO	    & 2021--08--12    & DOLORES \\
149	& 1395648	& 11:30:10.63	& --21:51:51.40	& 17.554	& 3.373	& QSO	    & 2021--04--20    & LDSS3 \\
150	& 1395793	& 11:51:30.64	& --02:42:42.99	& 17.524	& 2.605	& QSO	    & 2021--04--08    & DOLORES \\
151	& 1396709*	& 20:10:18.71	& --34:34:52.04	& 17.533	& 2.654	& QSO	    & 2021--07--25    & LDSS3 \\
152	& 1397485	& 22:09:08.00	& --46:34:49.28	& 17.570	& 2.434	& QSO	    & 2021--10--24    & LDSS3 \\
153	& 1398450	& 04:49:42.32	& --48:39:17.77	& 17.528	& 3.32	& QSO	    & 2021--09--22    & LDSS3 \\
154	& 1399299	& 03:58:56.29	& +04:07:51.57	& 17.568	& 2.819	& QSO	    & 2021--10--11    & DOLORES \\
155	& 1400749*	& 22:37:11.66	& --30:34:12.48	& 17.683	& 2.885	& QSO	    & 2021--07--24    & LDSS3 \\
156	& 1400778	& 22:34:43.75	& --25:58:57.76	& 17.650	& 3.82	& QSO	    & 2021--07--24    & LDSS3 \\
157	& 1401072	& 22:19:07.66	& --15:26:17.25	& 17.610	& 2.87	& QSO	    & 2021--08--13    & DOLORES \\
158	& 1401906	& 00:43:10.10	& --08:22:54.56	& 17.635	& 2.659	& QSO	    & 2021--08--13    & DOLORES \\
159	& 1403654	& 14:20:05.46	& --31:57:51.18	& 17.669	& 2.478	& QSO	    & 2021--04--16    & LDSS3 \\
160	& 1405092	& 23:59:12.73	& --65:39:58.31	& 17.608	& 2.946	& QSO	    & 2021--11--16    & EFOSC \\
161	& 1407972	& 17:00:12.13	& +05:23:11.15	& 17.626	& 0.0	& Star	    & 2021--04--17    & LDSS3 \\
162	& 1408011	& 17:18:47.53	& +09:58:05.61	& 17.640	& 0.0	& Star	    & 2021--04--17    & LDSS3 \\
163	& 1408772	& 23:40:29.20	& --17:14:56.45	& 17.705	& 2.666	& QSO	    & 2021--08--13    & DOLORES \\
164	& 1410714	& 06:06:56.47	& --41:57:24.60	& 17.709	& 2.44	& QSO	    & 2021--10--23    & LDSS3 \\
165	& 1411425	& 14:19:27.92	& --29:33:06.03	& 17.761	& 2.517	& QSO	    & 2021--06--01    & LDSS3 \\
166	& 1412590	& 18:36:02.62	& --75:15:13.50	& 17.736	& 2.592	& QSO	    & 2021--06--01    & LDSS3 \\
167	& 1412709	& 20:36:38.67	& --74:44:34.77	& 17.775	& 3.523	& QSO	    & 2021--07--25    & LDSS3 \\
168	& 1414388	& 21:40:12.05	& +07:38:56.57	& 17.752	& 0.0	& Star	    & 2021--06--18    & DOLORES \\
169	& 1414414*	& 22:26:00.20	& +03:10:45.26	& 17.736	& 2.573	& QSO	    & 2021--10--10    & DOLORES \\
170	& 1416050	& 21:39:03.54	& --17:28:45.49	& 17.812	& 1.844	& QSO	    & 2021--08--13    & DOLORES \\
171	& 1416355*	& 23:48:44.91	& --18:58:57.82	& 17.896	& 2.352	& QSO	    & 2021--08--13    & DOLORES \\
172	& 1417060*	& 01:43:26.02	& --22:19:06.67	& 17.805	& 2.866	& QSO	    & 2021--10--23    & LDSS3 \\
173	& 1417101	& 01:18:08.40	& --14:39:54.05	& 17.862	& 3.432	& QSO	    & 2021--10--11    & DOLORES \\
174	& 1417973	& 03:49:40.22	& --19:04:01.17	& 17.820	& 2.998	& QSO	    & 2021--10--28    & DOLORES \\
175	& 1418007	& 03:49:27.90	& --13:39:29.29	& 17.818	& 2.552	& QSO	    & 2021--10--28    & DOLORES \\
176	& 1418068	& 04:12:07.76	& --07:26:41.34	& 17.810	& 3.155	& QSO	    & 2021--10--28    & DOLORES \\
177	& 1418187*	& 05:35:35.58	& --22:43:18.87	& 17.830    & 2.781	& QSO	    & 2021--10--23    & LDSS3 \\
178	& 1418636*	& 11:40:44.94	& --14:00:06.95	& 17.802	& 2.372	& QSO	    & 2021--04--16    & DOLORES \\
179	& 1419028	& 14:18:22.49	& --29:50:16.23	& 17.828	& 2.621	& QSO	    & 2021--04--16    & LDSS3 \\
180	& 1419866	& 20:23:45.02	& --30:02:43.08	& 17.868	& 2.922	& QSO	    & 2021--07--27    & LDSS3 \\
181	& 1420234	& 20:00:44.85	& --52:57:01.67	& 17.874	& 3.431	& QSO	    & 2021--09--23    & LDSS3 \\
182	& 1420343	& 20:25:25.75	& --87:20:08.28	& 17.839	& 2.995	& QSO	    & 2021--10--23    & LDSS3 \\
183	& 1420637*	& 23:54:09.70	& --59:33:54.65	& 17.884	& 3.318	& QSO	    & 2021--11--14    & EFOSC \\
184	& 1421715	& 04:24:21.65	& --42:43:56.11	& 17.868	& 3.341	& QSO	    & 2021--09--22    & LDSS3 \\
185	& 1423061	& 16:38:36.96	& +06:21:14.57	& 17.890	& 0.408	& Galaxy    & 2021--06--18    & DOLORES \\
186	& 1424197*	& 01:13:39.52	& --32:00:39.88	& 17.960	& 2.873	& QSO	    & 2021--10--24    & LDSS3 \\
187	& 1424328	& 02:02:03.86	& --28:45:13.11	& 17.984	& 2.745	& QSO	    & 2021--10--24    & LDSS3 \\
188	& 1424706	& 00:57:22.23	& --12:28:39.67	& 17.990	& 3.242	& QSO	    & 2021--08--13    & DOLORES \\
189	& 1424813*	& 02:00:58.93	& --18:50:10.85	& 17.975	& 3.05	& QSO	    & 2021--08--13    & DOLORES \\
190	& 1425764	& 05:20:11.95	& --22:12:38.07	& 17.907	& 3.217	& QSO	    & 2021--10--24    & LDSS3 \\
191	& 1425771	& 05:40:40.99	& --25:37:35.79	& 17.982	& 3.541	& QSO	    & 2021--10--23    & LDSS3 \\
192	& 1425875	& 04:51:12.27	& --03:54:52.72	& 17.907	& 2.515	& QSO	    & 2021--10--28    & DOLORES \\
193	& 1426032	& 09:46:17.18	& --19:43:25.25	& 17.918	& 4.096	& QSO	    & 2021--11--16    & EFOSC \\
194	& 1426515	& 13:12:15.49	& --33:53:35.89	& 17.978	& 3.945	& QSO	    & 2021--04--16    & LDSS3 \\
195	& 1426540	& 13:17:41.33	& --31:29:30.01	& 17.949	& 3.128	& QSO	    & 2021--04--16    & LDSS3 \\
196	& 1427009	& 13:05:51.74	& --08:46:20.53	& 17.975	& 3.009	& QSO	    & 2021--06--17    & DOLORES \\
197	& 1427108	& 14:52:08.88	& --13:27:35.29	& 17.979	& 3.125	& QSO	    & 2021--04--08    & DOLORES \\
198	& 1427514	& 19:50:09.10	& --36:12:17.34	& 17.912	& 2.316	& QSO	    & 2021--06--01    & LDSS3 \\
199	& 1427938	& 19:46:00.82	& --54:33:54.24	& 17.999	& 3.376	& QSO	    & 2021--05--31    & LDSS3 \\
200	& 1427979	& 18:58:41.72	& --64:49:24.11	& 17.991	& 2.328	& QSO	    & 2021--06--01    & LDSS3 \\
201	& 1428039	& 19:33:00.37	& --72:19:35.63	& 17.972	& 3.084	& QSO	    & 2021--05--31    & LDSS3 \\
202	& 1429220*	& 02:14:28.51	& --59:11:43.94	& 17.987	& 3.956	& QSO	    & 2021--09--24    & LDSS3 \\
203	& 1429844	& 04:06:53.40	& --42:46:44.74	& 17.991	& 3.798	& QSO	    & 2021--11--15    & EFOSC \\
204	& 1430868	& 21:59:59.36	& --83:39:50.94	& 17.937	& 3.709	& QSO	    & 2021--07--24    & LDSS3 \\
205	& 1430911	& 20:55:20.45	& --55:28:15.22	& 17.912	& 3.427	& QSO	    & 2021--09--23    & LDSS3 \\
206	& 1431053	& 00:02:58.83	& --46:19:43.81	& 17.943	& 2.455	& QSO	    & 2021--10--24    & LDSS3 \\
\bottomrule
\end{longtable}





%


\bsp	
\label{lastpage}
\end{document}